\documentclass{aa}

\usepackage{graphicx}
\usepackage{xcolor}
\usepackage{txfonts}
\usepackage{subfigure}
\usepackage{float}
\usepackage{natbib,twoopt}
\usepackage{xspace}
\usepackage{balance}
\usepackage{upgreek}
\usepackage{multirow}
\usepackage[normalem]{ulem}
\usepackage{soul}
\usepackage[colorlinks=true, citecolor=blue]{hyperref}

\usepackage[flushleft]{threeparttable}

\definecolor{redUniBo}{RGB}{187, 46, 41}

\def\apjl{ApJ Lett.}
\def\apjs{ApJ Suppl.}

\def\aap{A\&A}

\newcommand\dd{\mathord{\mathrm{d}}}
\newcommand{\Lagr}{\mathcal{L}}

\bibpunct{(}{)}{;}{a}{}{,} 
\DeclareGraphicsExtensions{.eps,.ps,.pdf}

\begin{document}

\title{AMICO galaxy clusters in KiDS-DR3: cosmological constraints from counts and stacked weak-lensing}

\author
{G. F. Lesci\inst{\ref{1},\ref{2}} 
\and F. Marulli\inst{\ref{1},\ref{2},\ref{3}}
\and L. Moscardini\inst{\ref{1},\ref{2},\ref{3}}
\and M. Sereno\inst{\ref{2},\ref{3}}
\and A. Veropalumbo\inst{\ref{4}}
\and M. Maturi\inst{\ref{5},\ref{8}}
\and C. Giocoli\inst{\ref{1},\ref{2},\ref{3}}
\and \\M. Radovich\inst{\ref{7}}
\and F. Bellagamba\inst{\ref{1},\ref{2}}
\and M. Roncarelli\inst{\ref{1},\ref{2}}
\and S. Bardelli\inst{\ref{2}}
\and S. Contarini\inst{\ref{1},\ref{2},\ref{3}}
\and G. Covone\inst{\ref{9},\ref{10},\ref{11}}
\and L. Ingoglia\inst{\ref{9}}
\and \\L. Nanni\inst{\ref{6},\ref{1}}
\and E. Puddu\inst{\ref{9}}
}

\offprints{G. F. Lesci \\ \email{giorgio.lesci@studio.unibo.it}}

\institute{
  Dipartimento di Fisica e Astronomia ``Augusto Righi'' - Alma Mater Studiorum
  Universit\`{a} di Bologna, via Piero Gobetti 93/2, I-40129 Bologna,
  Italy\label{1}
  \and INAF - Osservatorio di Astrofisica e Scienza dello Spazio di
  Bologna, via Piero Gobetti 93/3, I-40129 Bologna, Italy\label{2}
  \and INFN - Sezione di Bologna, viale Berti Pichat 6/2, I-40127
  Bologna, Italy\label{3}
  \and Dipartimento di Fisica, Universit\`{a} degli Studi Roma Tre, via
  della Vasca Navale 84, I-00146 Roma, Italy\label{4}
  \and Zentrum f\"ur Astronomie, Universit\"at Heidelberg, Philosophenweg 12, D-69120 Heidelberg, Germany\label{5}
  \and ITP, Universit\"at Heidelberg, Philosophenweg 16, D-69120 Heidelberg, Germany\label{8}
  \and INAF - Osservatorio Astronomico di Padova, vicolo dell'Osservatorio 5, I-35122 Padova, Italy\label{7}
  \and INAF - Osservatorio Astronomico di Capodimonte, Salita Moiariello 16, I-80131 Napoli, Italy\label{9}
  \and Dip. di Fisica “E. Pancini”, Universit\`{a} di Napoli Federico II, C.U. di Monte Sant’Angelo, Via Cintia, I-80126 Napoli, Italy\label{10}
  \and INFN, Sez. di Napoli, via Cintia, I-80126 Napoli, Italy\label{11}
  \and Institute of Cosmology \& Gravitation, University of Portsmouth, Dennis Sciama Building, Portsmouth, PO1 3FX, UK\label{6}
}

\date{Received --; accepted --}

\newcommand\omegam{0.24}
\newcommand\omegamUp{0.03}
\newcommand\omegamLow{0.04}
\newcommand\sigmaotto{0.86}
\newcommand\sigmaUp{0.07}
\newcommand\sigmaLow{0.07}
\newcommand\Sotto{0.78}
\newcommand\SUp{0.04}
\newcommand\SLow{0.04}
\newcommand\ascal{0.04}
\newcommand\aup{0.04}
\newcommand\alow{0.03}
\newcommand\bscal{1.72}
\newcommand\bup{0.08}
\newcommand\blow{0.08}
\newcommand\cscal{-2.37}
\newcommand\cscalup{0.37}
\newcommand\clow{0.56}
\newcommand\scatter{0.18}
\newcommand\scatterUp{0.08}
\newcommand\scatterLow{0.10}
\newcommand\scatterM{0.11}
\newcommand\scatterMUp{0.19}
\newcommand\scatterMLow{0.22}

\abstract
   {}
   {We present a cosmological analysis of abundances and stacked weak-lensing profiles of galaxy clusters, exploiting the AMICO KiDS-DR3 catalogue. The sample consists of 3652 galaxy clusters with intrinsic richness $\lambda^*\geq20$, over an effective area of 377\,deg$^2$, in the redshift range $z\in[0.1,\,0.6]$.
   }
   {We quantified the purity and completeness of the sample through simulations. The statistical analysis has been performed by simultaneously modelling the comoving number density of galaxy clusters and the scaling relation between the intrinsic richnesses and the cluster masses, assessed through a stacked weak-lensing profile modelling. The fluctuations of the matter background density, caused by super-survey modes, have been taken into account in the likelihood. Assuming a flat $\Lambda$CDM model, we constrained $\Omega_{\rm m}$, $\sigma_8$, $S_8 \equiv \sigma_8(\Omega_{\rm m}/0.3)^{0.5}$, and the parameters of the mass-richness scaling relation.}
   {We obtained $\Omega_{\rm m}=\omegam^{+\omegamUp}_{-\omegamLow}$, $\sigma_8=\sigmaotto^{+\sigmaUp}_{-\sigmaLow}$, $S_8=\Sotto^{+\SUp}_{-\SLow}$. The constraint on $S_8$ is consistent within 1$\sigma$ with the results from WMAP and Planck. 
   Furthermore, we got constraints on the cluster mass scaling relation in agreement with those obtained from a previous weak-lensing only analysis.
   }
   {}

\keywords{clusters -- Cosmology: observations -- large-scale structure
  of Universe -- cosmological parameters}

\authorrunning{G. F. Lesci et al.}

\titlerunning{Cosmological constraints from AMICO KiDS-DR3 cluster
  counts and weak-lensing}

\maketitle

\section{Introduction}
Galaxy clusters lie at the nodes of the cosmic web, tracing the deepest virialised potential wells of the dark matter distribution in the present Universe.
Large samples of galaxy clusters can be built by exploiting different
techniques, thanks to their multi-wavelength emission. In particular, galaxy clusters can be detected through the bremsstrahlung emission of the intracluster medium in the X-ray band
\citep[e.g.][]{boringer, clerc, pierre2016}, through the detection of
the Sunyaev-Zel’dovich effect in the cosmic microwave background (CMB)
\citep[e.g.][]{hilton}, or through their gravitational lensing effect on the
background galaxies \citep[e.g.][]{maturicit,satoshi}. Furthermore, galaxy clusters can
be detected in the optical band by looking  for overdensities and peculiar features characterising cluster members in 
galaxy surveys
\citep[e.g.][]{redmapper, amico}.

The number counts and clustering of galaxy clusters are effective probes to constrain the geometrical and dynamical properties of the Universe \citep[see e.g.][and references therein]{vikhlinin2009, veropalumbo2014, veropalumbo2016, sereno2015, marulli2017, marulli2018, marulli2020, pacaud, costanzi, nanni2020}. The formation and evolution of galaxy clusters, being mostly driven by gravity, can be followed with high accuracy using N-body simulations \citep{borgani2011, angulo2012, giocoli1}, that allow in particular to calibrate, from a theoretical point of view, the functional form of the dark matter halo mass function in different cosmological scenarios \citep[e.g.][]{shethtormen, tinker, watson, despali}. Many attempts have been also made to investigate the impact of the baryonic physical processes on cluster statistics, including the mass function \citep[e.g.][]{cui2012, velliscig, bocquet, castro2020}. In addition, promising techniques to constrain the cosmological parameters concern the study of the weak-lensing peak counts in cosmic shear maps \citep[e.g.][]{peaks2,peaks1,shan17,martinet17,giocoliPeaks} and galaxy cluster sparsity \citep[e.g.][]{balmes14,corasaniti21}. \\
\indent Although it is possible to predict with great accuracy the abundance of dark matter haloes as a function of mass for a given cosmological model, the cluster masses cannot be easily derived from observational data. Currently, the most reliable mass measurements are provided by weak gravitational lensing \citep[e.g.][]{citlens1,citlens2,citlens3,citlens4,citlens5,citlens6}, that consists in the deflection of the light rays coming from background sources, due to the intervening cluster potential, and it accounts for both the dark and baryonic matter components. Differently from the other methods to estimate cluster masses based on the properties of the gas and of the member galaxies, such as the ones exploiting X-ray emission, galaxy velocity dispersion, or the Sunyaev-Zel’dovich effect on the CMB, the gravitational lensing method does not rely on any assumption on the dynamical state of the cluster. However, weak-lensing mass measurements of individual galaxy clusters are feasible only if the  signal-to-noise ratio (S/N) of the shear profiles is sufficiently high: this requires either
a massive structure or deep observations. Thus, in cosmological studies of cluster statistics, it is often necessary to stack the weak-lensing signal produced by a set of objects with similar properties, from which a mean value of their mass is estimated. These mean mass values can be linked to a direct observable, or mass proxy, which can be used to define a mass-observable scaling relation. Such a mass proxy could be e.g.\ the luminosity, the pressure or the temperature of the intracluster medium, in the case of X-ray observations, the richness (i.e.\ the number of member galaxies) for optical surveys, or the pressure measured from the cluster Sunyaev-Zel'dovich signal. \\
\indent In addition to the mass scaling relation, a crucial quantity that has to be estimated accurately in any cosmological study of clusters is the selection function of the sample. In fact, it is crucial to properly account for purity and completeness of the cluster catalogue. Once the selection function and the mass-observable scaling relation have been accurately assessed, galaxy cluster statistics provide
powerful cosmological constraints in the low-redshift Universe, which can be combined with high-redshift constraints from the
CMB. In
particular, by measuring the abundance of clusters as a function of
mass and redshift it is possible to provide constraints on the matter density
parameter, $\Omega_{\rm m}$, and on the amplitude of the matter power spectrum,
$\sigma_8$. The constraints on $\sigma_8$ from clusters can be
combined with the primordial matter power spectrum amplitude
constrained by the CMB, in order to assess the growth rate of cosmic structures. Moreover, combining cluster statistics with distance
measurements, e.g. from Baryon Acoustic Oscillations or Type Ia supernovae data, provides constraints on the total energy density of
massive neutrinos, $\Omega_\nu$, on the normalised Hubble constant, $h\equiv H_0/(100$ km/s/Mpc), and on the dark energy equation of state parameter, $w$
\citep{allen}. Ongoing wide extra-galactic surveys, such as the Kilo Degree Survey
(KiDS)\footnote{\url{http://kids.strw.leidenuniv.nl/}}, the Dark Energy
Survey\footnote{\url{ https://www.darkenergysurvey.org}}
\citep{desCollab}, the surveys performed with the South Pole Telescope\footnote{\url{https://pole.uchicago.edu/}} and with the Atacama Cosmology Telescope\footnote{\url{ https://act.princeton.edu/}}, and future projects, such as Euclid\footnote{\url{http://sci.esa.int/euclid/}} \citep{laureijs, sartoriseuclid, amendola2018}, the Vera C. Rubin Observatory LSST\footnote{\url{ https://www.lsst.org/}} \citep{LSST2012},
eRosita\footnote{\url{ http://www.mpe.mpg.de/eROSITA}}, and the Simons Observatory survey\footnote{\url{ https://simonsobservcatory.org/}} will provide highly complete and pure cluster catalogues up to high redshifts and low masses. \\
\begin{figure}[t]
\centering \includegraphics[height=5.8cm,width=\hsize]{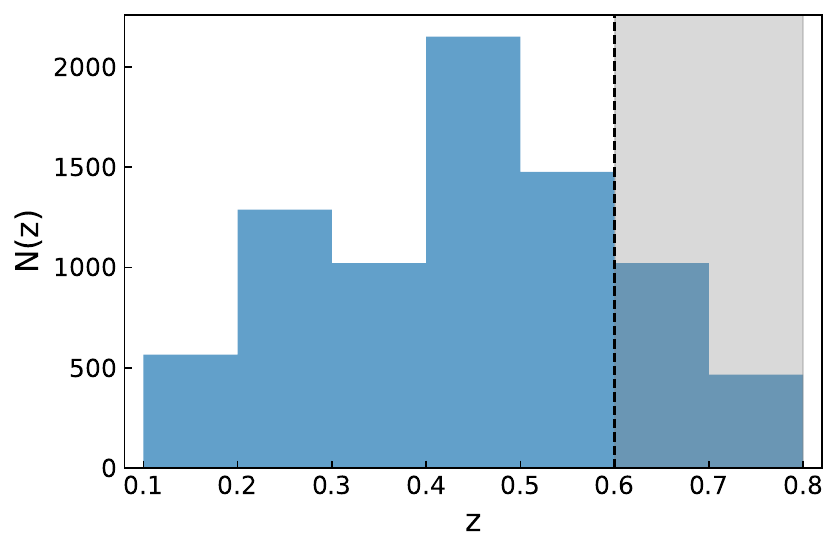}
\caption{Distribution of the clusters as a function of redshift. Objects with $z>0.6$, not considered in the cosmological analysis, are covered by the shaded grey area.}
\label{reddistr}
\end{figure}
\indent In this work we analyse a catalogue of 3652 galaxy clusters \citep{catalogue} identified with the Adaptive Matched Identifier of Clustered Objects (AMICO) algorithm \citep{amico} in the third data release of the Kilo Degree Survey \citep[KiDS-DR3;][]{dejong}. We measure the number counts of the clusters in the sample as a function of the intrinsic richness, $\lambda^*$, used as a mass proxy, and of the redshift, $z$. In addition, we estimate the mean values of cluster masses in bins of $\lambda^*$ and $z$, following a stacked weak-lensing analysis as in \citet{lensing}. Then we model simultaneously the cluster number counts and the masses, through a Bayesian approach consisting in a Markov chain Monte Carlo (MCMC) analysis, taking into account the selection function of the sample. Assuming a flat $\Lambda$ cold dark matter ($\Lambda$CDM) framework, we constrain the cosmological parameters $\sigma_8$, $\Omega_{\rm m}$, $S_8\equiv \sigma_8(\Omega_{\rm m}/0.3)^{0.5}$, and the scaling relation parameters, including its intrinsic scatter, $\sigma_{\rm intr}$. The analysis of cluster clustering within this dataset is performed in \citet{nanni2020}, where we obtain results in agreement with those presented in this work. \\
\indent The whole cosmological analysis is performed using the {\small CosmoBolognaLib}\footnote{\url{https://gitlab.com/federicomarulli/CosmoBolognaLib/}} (CBL) \citep{cbl} V5.4, a large set of {\em free software} C++/Python libraries that provide an efficient numerical environment for cosmological investigations of the
large-scale structure of the Universe.\\
\indent The paper is organised as follows. In Section \ref{dataset} we present the AMICO KiDS-DR3 cluster catalogue and the weak-lensing dataset, introducing also the mass-observable scaling relation. In addition, we discuss the methods considered in this work to estimate the selection function of the sample and the uncertainties related to the cluster properties. In Section \ref{SectionMethods} we present the theoretical model used to describe the galaxy cluster counts, along with the likelihood function. The results are presented and discussed in Section \ref{SecResults}, leading to our conclusions summarised in Section \ref{SecConclusions}.


\begin{table*}[t]
\caption{\label{tab2}Cluster binning used for the weak-lensing analysis.}
  \centering
    \begin{tabular}{c c c c c c c} 
      $z$ range & $z_{\text{eff}}$ & $\lambda^*$ range & $\lambda^*_{\text{eff}}$ & $\log\,[\bar{M}_{200}/(10^{14}\,\text{M$_\odot$}\,h^{-1})]$ & $N_{\text{cl}}$ & $z_{\text{s,eff}}$\\
      \hline
      \rule{0pt}{4ex}
	  $[0.10,\,0.30]$ & $0.189\pm0.001$ & $[0,\,15]$ & $10.20\pm0.09$ & -0.73\,$\pm$\,0.07 & 1246 & $0.849\pm0.002$ \\\rule{0pt}{2.5ex}
     $[0.10,\,0.30]$ & $0.212\pm0.002$ & $[15,\,25]$ & $18.88\pm0.12$ & -0.38\,$\pm$\,0.07 & 684 & $0.867\pm0.002$ \\\rule{0pt}{2.5ex}
     $[0.10,\,0.30]$ & $0.222\pm0.004$ & $[25,\,35]$ & $29.02\pm0.21$ & 0.05\,$\pm$\,0.07 & 209 & $0.879\pm0.002$ \\\rule{0pt}{2.5ex}
     $[0.10,\,0.30]$ & $0.228\pm0.007$ & $[35,\,45]$ & $39.75\pm0.32$ & 0.32\,$\pm$\,0.06 & 82 & $0.877\pm0.005$ \\\rule{0pt}{2.5ex}
     $[0.10,\,0.30]$ & $0.222\pm0.008$ & $[45,\,150]$ & $56.59\pm2.20$ & 0.54\,$\pm$\,0.06 & 44 & $0.890\pm0.013$ \\\rule{0pt}{4ex}
     $[0.30,\,0.45]$ & $0.374\pm0.001$ & $[0,\,20]$ & $15.10\pm0.11$ & -0.41\,$\pm$\,0.08 & 1113 & $0.948\pm0.003$ \\\rule{0pt}{2.5ex}
     $[0.30,\,0.45]$ & $0.387\pm0.002$ & $[20,\,30]$ & $24.08\pm0.11$ & -0.07\,$\pm$\,0.07 & 767 & $0.944\pm0.004$ \\\rule{0pt}{2.5ex}
     $[0.30,\,0.45]$ & $0.389\pm0.002$ & $[30,\,45]$ & $35.91\pm0.27$ & 0.21\,$\pm$\,0.06 & 320 & $0.941\pm0.005$ \\\rule{0pt}{2.5ex}
     $[0.30,\,0.45]$ & $0.390\pm0.005$ & $[45,\,60]$ & $50.88\pm0.50$ & 0.41\,$\pm$\,0.08 & 87 & $0.950\pm0.015$ \\\rule{0pt}{2.5ex}
     $[0.30,\,0.45]$ & $0.379\pm0.006$ & $[60,\,150]$ & $73.60\pm2.09$ & 0.68\,$\pm$\,0.07 & 45 & $0.946\pm0.012$ \\\rule{0pt}{4ex}
     $[0.45,\,0.60]$ & $0.498\pm0.001$ & $[0,\,25]$ & $19.71\pm0.11$ & -0.33\,$\pm$\,0.09 & 1108 & $0.958\pm0.001$ \\\rule{0pt}{2.5ex}
     $[0.45,\,0.60]$ & $0.514\pm0.002$ & $[25,\,35]$ & $29.23\pm0.12$ & -0.07\,$\pm$\,0.07 & 761 & $0.961\pm0.004$ \\\rule{0pt}{2.5ex}
     $[0.45,\,0.60]$ & $0.523\pm0.003$ & $[35,\,45]$ & $39.25\pm0.18$ & 0.21\,$\pm$\,0.07 & 299 & $0.961\pm0.006$ \\\rule{0pt}{2.5ex}
     $[0.45,\,0.60]$ & $0.513\pm0.004$ & $[45,\,150]$ & $55.12\pm0.76$ & 0.36\,$\pm$\,0.07 & 197 & $0.960\pm0.004$
    \end{tabular}
    \tablefoot{The computation of $z_{\text{eff}}$ and $\lambda^*_{\text{eff}}$ and their uncertainties is described in \citet{lensing}. For the logarithm of the measured mean masses, $\log\bar{M}_{200}$, we quote the mean and the standard deviation of the posterior probability distribution. $N_{\text{cl}}$ is the number of clusters in the bin. In the last column, $z_{\text{s,eff}}$ is the effective redshift of the lensed sources, obtained by following the procedure described in \citet{giocoli2020}. Quoted masses refer to a flat $\Lambda$CDM model with $\Omega_{\rm m}=0.3$ and $h=0.7$.}
\end{table*}

\section{Dataset}\label{dataset}

\subsection{The catalogue of galaxy clusters}
The catalogue of galaxy clusters this work is based on, named AMICO KiDS-DR3 \citep{catalogue}, is derived from the third data release of the Kilo Degree Survey \citep{dejong}. The KiDS survey has been carried out with the OmegaCAM wide-field imager \citep{Kuijken} mounted at the VLT Survey Telescope, a 2.6 m telescope sited at the Paranal Observatory. In particular, the 2\,arcsec aperture photometry in \textit{u}, \textit{g}, \textit{r}, \textit{i} bands is provided, as well as the photometric redshifts for all galaxies down to the 5$\sigma$ limiting magnitudes of 24.3, 25.1, 24.9 and 23.8 for the aforementioned four bands, respectively. For the final galaxy cluster catalogue considered for cluster counts \citep{catalogue}, only the galaxies with magnitude $r<24$ have been considered, for a total of about 32 million galaxies. For the weak-lensing analysis, instead, no limits in magnitude have been imposed for the lensed sources in order to exploit the whole dataset available, and the catalogue \citep[developed by][]{hildebr} provides the shear measurements for about 15 million galaxies. \\
\indent Galaxy clusters have been detected thanks to the application of the AMICO algorithm \citep{amico}, which identifies galaxy overdensities by exploiting a linear matched optimal filter. In particular, the detection process adopted for this study relies solely on angular coordinates, magnitudes and photometric redshifts (photo-$z$s from now on) of galaxies. 
Unlike other algorithms used in literature for cluster identification, AMICO does not use any direct information coming from colours, like for instance the so-called red sequence. For this reason, AMICO is expected to be accurate also at higher redshifts, where the red sequence may not yet be prominent. The excellent performances of AMICO have been recently confirmed by the analysis made in \citet{adam2019}, where the purity and completeness of the cluster catalogues extracted by applying six different algorithms on realistic mock catalogues reproducing the expected characteristics of the future Euclid photometric survey \citep{laureijs}
have been compared. As a result of this challenge, AMICO is one of the two algorithms for cluster identification officially adopted by the Euclid mission. \\
\indent The KiDS-DR3 sample covers a total area of 438 deg$^2$, but all the galaxies located in the regions affected by image artifacts, or falling in the secondary/tertiary halo masks used for the weak-lensing analysis \citep[see][]{dejong2015}, have been rejected. This yields a final effective area of 377 deg$^2$, containing all the cluster detections with $S/N>3.5$ and within the redshift range $z\in[0.1,0.8]$, for a total of 7988 objects. Due to the low signal-to-noise ratio of the shear profiles for $z>0.6$, not sufficient to perform a stacked weak-lensing analysis, we decided to exclude the redshift bin $z\in[0.6,0.8]$ from the analysis. In Fig. \ref{reddistr} we show the redshift distribution of the AMICO KiDS-DR3 cluster sample. 

\subsection{Mass proxy}
We exploit the cluster shear signal through a stacked weak-lensing analysis to estimate the mean cluster masses in bins of intrinsic richness and redshift. The intrinsic richness, $\lambda^*$, is defined as:
\begin{equation}\label{lambda}
\lambda^*_j=\sum\limits_{i=1}^{N_{\rm gal}} P_i(j)\;\;\;\;\text{with}\;\;\;\;
\begin{cases}
m_i<m^*(z_j)+1.5 \\ R_i(j)<{R_{\rm max}(z_j)}
\end{cases}
,
\end{equation}
where $P_i(j)$ is the probability, assigned by AMICO, that the $i$-th galaxy is a member of a given detection $j$ \citep[see][]{catalogue}. The intrinsic richness thus represents the sum of the membership probabilities, that is the weighted number of visible galaxies belonging to a detection, under the conditions given by Eq.\ \eqref{lambda}. The sum of the membership probabilities is an excellent estimator of the true number of member galaxies, as shown in \citet{amico} by running the AMICO algorithm on mock catalogues (see Fig.\ 8 in the reference). In particular, in Eq.\ \eqref{lambda}, $z_j$ is the redshift of the $j$-th detected cluster, $m_i$ is the magnitude of the $i$-th galaxy, and $R_i$ corresponds to the distance of the $i$-th galaxy from the centre of the cluster. The parameter $R_{\rm max}(z_j)$ represents the radius enclosing a mass $M_{200}=10^{14}M_\odot/h$, such that the corresponding mean density is 200 times the critical density of the Universe at the given redshift $z_j$. In the following analysis, indeed, we consider the masses evaluated as $M_{200}$. Lastly, $m^*$ is the typical magnitude of the Schechter function in the cluster model assumed in the AMICO algorithm. We use the term \textit{intrinsic richness} as opposed to the \textit{apparent richness}, defined in \citet{catalogue}. In particular, since the threshold in absolute magnitude is always lower than the survey limit thanks to its redshift dependence, $\lambda^*$ does not depend on the survey limit. Conversely, the apparent richness is a quantity that includes all visible galaxies and is therefore related to how a cluster is observed given a certain apparent magnitude limit. \\
\indent We set $\lambda^*=20$ as the threshold for the counts analysis, in order to exclude the bins affected by detection impurities and severe incompleteness. Thus the final sample of galaxy clusters considered for the counts analysis contains 3652 objects, with $\lambda^*\geq20$, in the redshift bins $z\in[0.1,0.3]$, $z\in[0.3,0.45]$, $z\in[0.45,0.6]$. With regard to the binning in intrinsic richness, we adopt 4 logarithmically spaced bins in the range $\lambda^*\in[20,\,137]$ for each redshift bin. To test the robustness of our results with respect to this binning choice, we repeated the analysis assuming different numbers of $\lambda^*$ bins obtaining negligible differences in the final results, i.e. much below the 1$\sigma$ of the posterior distributions, and values of reduced $\chi^2$ always consistent with 1. On the other hand, as we will discuss in the next section, in order to fully exploit the available data we do not impose any threshold in $\lambda^*$ in the weak-lensing masses analysis, and we choose a different binning in $\lambda^*$. To test the reliability of this approach, we also perform the cosmological analysis by imposing $\lambda^*\geq20$ for the weak-lensing masses, deriving results fully in agreement with those obtained without assuming this threshold, as detailed in Section \ref{SecResults}.

\subsection{Weak-lensing masses}\label{SectionLensing}
\begin{figure}[t]
   \centering
   \includegraphics[width=\hsize,height=\dimexpr\textheight-196pt\relax]{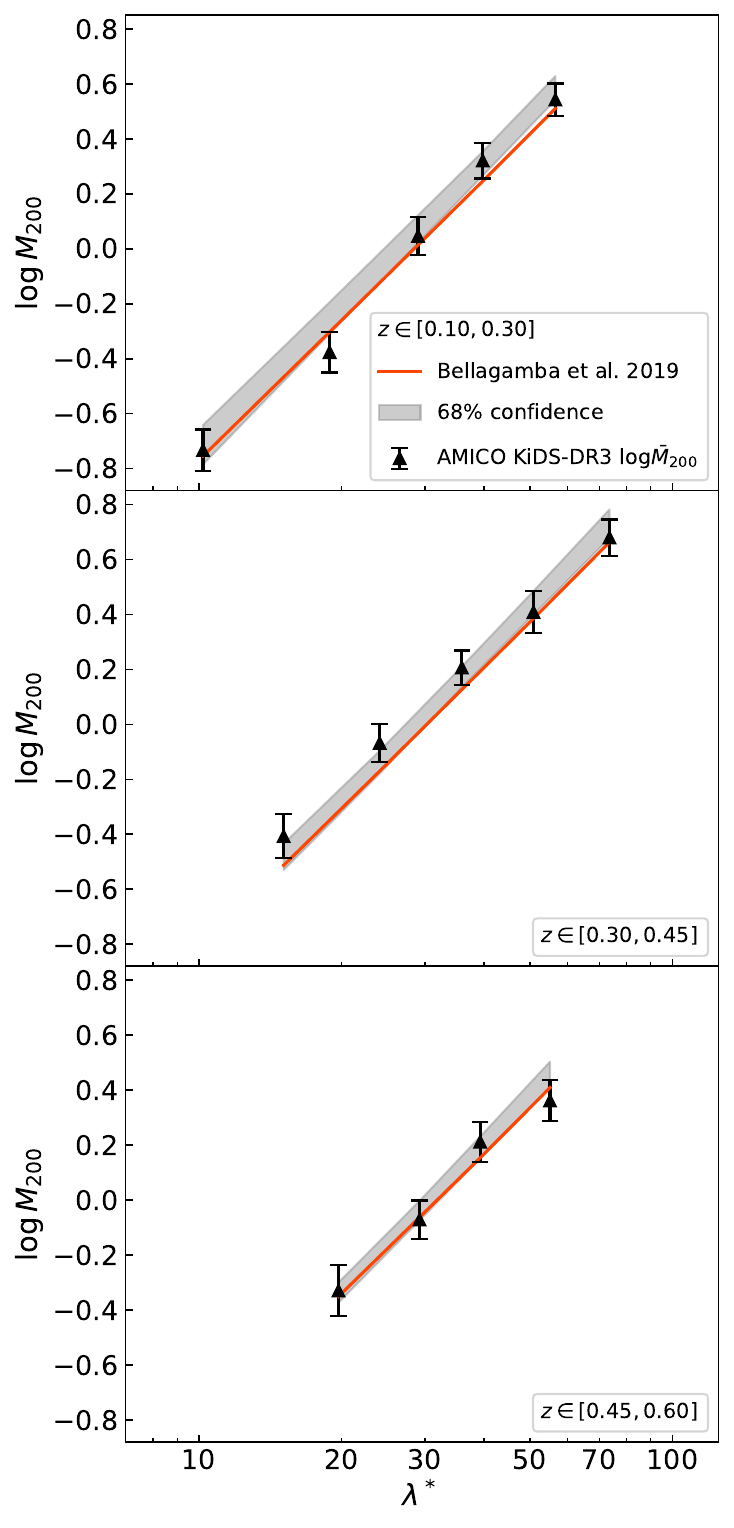}
      \caption{The logarithm of the masses in units of $(10^{14}\,\text{M$_\odot$}\,h^{-1})$, $\log\bar{M}_{200}$, from the AMICO KiDS-DR3 cluster catalogue as a
        function of the intrinsic richness $\lambda^*$, in the
        redshift bins $z\in[0.10,\,0.30]$, $z\in[0.30,\,0.45]$, $z\in[0.45,\,0.60]$, from top to bottom. The black triangles represent the mean values of $\log\bar{M}_{200}$, given by the mean of the marginalised posterior obtained in the weak-lensing analysis, while the error bars are given by 1$\sigma$ of the posterior distribution. The orange lines represent the median scaling relation obtained by modelling only $\log\bar{M}_{200}$, following the procedure described in \citet{lensing}. The grey bands represent the 68\% confidence level derived from the multivariate posterior of all the free parameters considered in the cosmological analysis described in Section \ref{SectionMethods}.}
         \label{figMasses}
   \end{figure}
\noindent To estimate the mean masses of the observed galaxy clusters we follow the same stacked weak-lensing procedure described in \citet{lensing}, based on KiDS-450 data. The clusters selected for the weak-lensing analysis lie in the redshift range $z\in[0.1,0.6]$, over an effective area of 360.3 deg$^2$. This area is slightly smaller compared to that considered for the counts analysis, since it derives from the masking described in \citet{hildebr}. Despite the availability of galaxy clusters up to $z=0.8$ in the AMICO KiDS-DR3 catalogue, the $S/N$ of the stacked shear profiles is too low to perform the stacking for $z>0.6$. Therefore we base our analysis on the redshift bins $z\in[0.1,0.3]$, $z\in[0.3,0.45]$, $z\in[0.45,0.6]$, deriving the estimated mean masses in a flat $\Lambda$CDM cosmology with $\Omega_{\rm m}=0.3$ and $h=0.7$. \\
\begin{figure*}[t]
\centering
\includegraphics[width = 0.4 \hsize, height = 6.8cm] {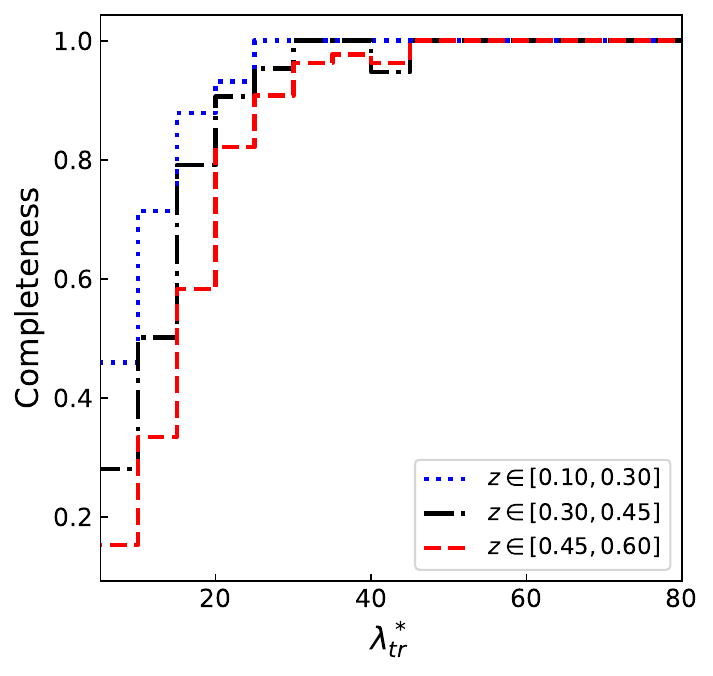}
\includegraphics[width = 0.4 \hsize, height = 6.8cm] {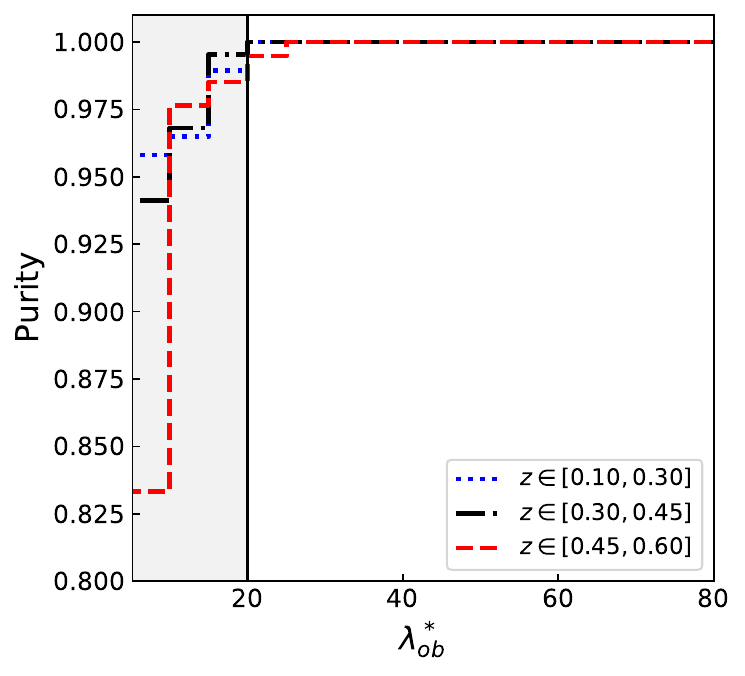}
\caption{The completeness (left panel) and the purity (right panel) of the AMICO KiDS-DR3 cluster catalogue as a function of the redshift, $z$, and of the intrinsic richness, $\lambda^*$. The completeness is a function of the true $\lambda^*$, i.e.\ $\lambda^*_{\rm tr}$, while the purity is defined as a function of the observed intrinsic richness, $\lambda^*_{\rm ob}$. The shaded area in the right panel highlights the bins excluded from the number counts analysis, i.e.\ for $\lambda^*_{\rm ob}<20$.}
\label{selfunc}
\end{figure*}
\indent With an MCMC analysis we sample the posterior distributions of the base 10 logarithm of the estimated mean cluster masses, $\log\bar{M}_{200}$, in 14 bins of intrinsic richness and redshift, considering $\lambda^*\geq0$, for a total of 6962 objects (see Table \ref{tab2}). As detailed in Section \ref{seclikelihood}, we account for the systematic errors affecting the weak-lensing mass estimates by relying on the results found in \citet{lensing}. Specifically, we consider the systematics due to background selection, photo-$z$s and shear measurements, affecting the measured stacked cluster profiles. Such errors are then propagated into the mass estimates. In particular, the sum in quadrature of such contributions to systematic errors, along with those produced by the halo model, orientation and projections, is equal to 7.6\%. The description of the modelling, including a more extensive discussion on the statistical and systematic uncertainties, is detailed in \citet{lensing}. \\
\indent The $\log\bar{M}_{200}$ posteriors are marginalised over the other parameters entering the modelling, that is the concentration parameter, $c_{200}$, the fraction of haloes belonging to the miscentred population, $f_{\rm off}$, and the rms of the distribution of the halo misplacement on the plane of the sky, $\sigma_{\rm off}$. In particular, we derive the posteriors for $c_{200}$, $f_{\rm off}$, $\sigma_{\rm off}$ in each bin, assuming the following flat priors: $c_{200}\in[1,\,20]$, $f_{\rm off}\in[0,\,0.5]$, $\sigma_{\rm off}\in[0\,{\rm Mpc}\,h^{-1},\,0.5\,{\rm Mpc}\,h^{-1}]$. Such parameters are not constrained by the data, that is their posteriors are statistically consistent with the priors. For what concerns the miscentering parameters, they are related to the possible difference between the center defined by the AMICO algorithm using the galaxy overdensities and the mass center related to the weak lensing signal. The uncertainty due to the use of a grid in AMICO impacts indeed only small scales not used in this analysis and then can be neglected. \\
\indent The logarithm of the estimated mean mass values, for different bins of intrinsic richness and redshift, are listed in Table \ref{tab2}. In Fig.\ \ref{figMasses} we show the median value of the mass-intrinsic richness scaling relation, described in Section \ref{sectionLensingModelling}, obtained by performing the modelling of the weak-lensing masses only as in \citet{lensing}, along with the 68\% confidence level obtained from the analysis of cluster counts and weak-lensing masses, detailed in Section \ref{SectionMethods}. It turns out that the $\log{M}_{200}-\log\lambda^*$ relation is reasonably linear and with an intrinsic scatter of $\sim 0.1$, as we will discuss in Section \ref{SecResults}, indicating the reliability of $\lambda^*$ as a mass proxy.

\subsection{Selection function}\label{SectionMock}
\begin{figure*}[t]
   \centering
   \includegraphics[width=\hsize,height=6.cm]{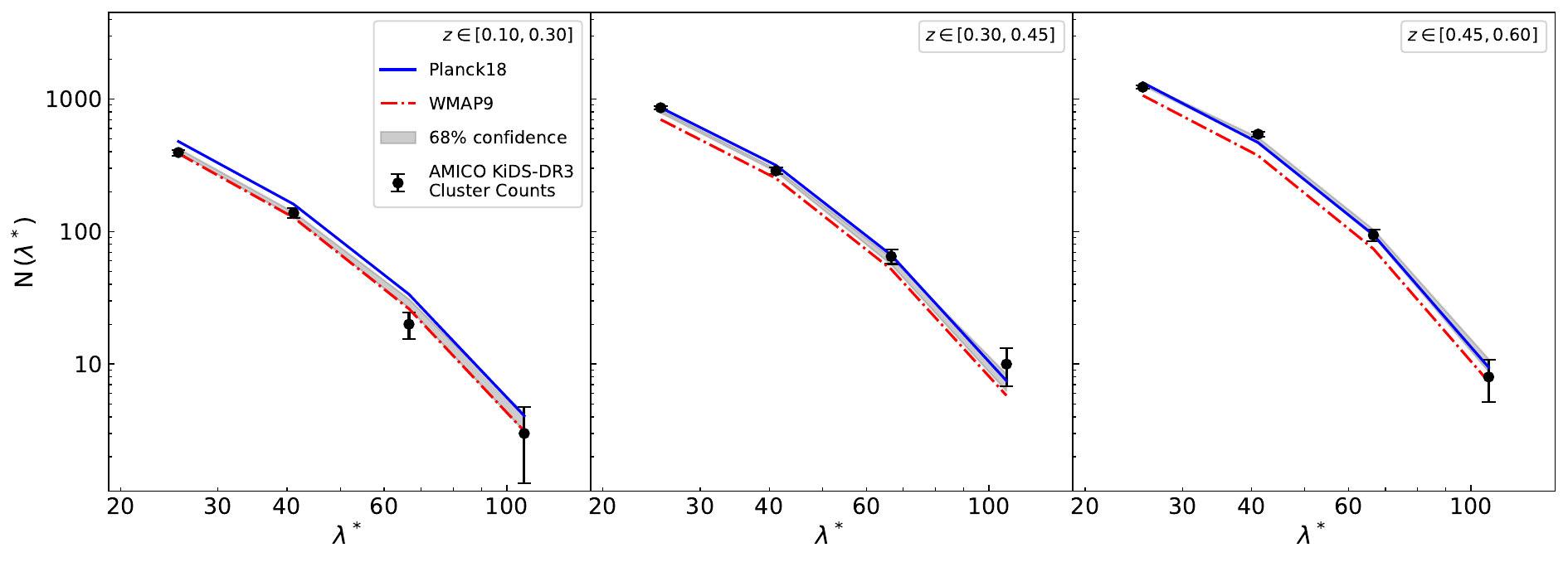}
      \caption{Number counts from the AMICO KiDS-DR3 cluster catalogue as a
        function of the intrinsic richness $\lambda^*$, in the
        redshift bins $z\in[0.10,\,0.30]$, $z\in[0.30,\,0.45]$, $z\in[0.45,\,0.60]$, from left to right. The black dots represent the counts directly retrieved from the catalogue, where the error bars are given by the Poissonian noise. The blue solid lines represent the model computed by assuming the cosmological parameters obtained by \citet{planck} (Table 2, TT,TE,EE+lowE), while the red dashed lines show the results based on the WMAP cosmological parameters \citep{wmap} (Table 3, WMAP-only Nine-year). Both in the Planck and WMAP cases, the scaling relation parameters and the intrinsic scatter have been fixed to the median values listed in Table \ref{tab1}, retrieved from the modelling. The grey bands represent the 68\% confidence level derived from the multivariate posterior of all the free parameters considered in the cosmological analysis.}
         \label{figCounts}
   \end{figure*}
In order to estimate the selection function of the AMICO KiDS-DR3 cluster catalogue, we make use of the mock catalogue described in \citet{catalogue}. The construction of the mock clusters is based on the original galaxy dataset, thus all the properties of the survey are properly taken into account, such as masks, photo-$z$ uncertainties and the clustering of galaxies. In this way the assumptions necessary to build up the mock catalogue are minimised. In particular, regarding the photo-$z$ uncertainties, the galaxies are drawn from the survey sample and selected from bins of richness and redshift, using a Monte Carlo sampling based on the cluster membership probability. The probability that  each galaxy is included in a given redshift bin is driven by its own photo-$z$ probability distribution function, which includes the contribution of the photometric noise. In this way, the selection of the simulated cluster member galaxies mimics the real uncertainties of the photometric redshifts in the photometric catalogue. Then, to derive the selection function, the AMICO code has been run on the mock catalogue, consisting of 9018 clusters distributed over a total area of 189 deg$^2$. Only the detections with $S/N>3.5$ are considered, this being the threshold applied to the real dataset. \\
\indent In Fig.\ \ref{selfunc} we show the purity and completeness of the dataset, which define the selection function. The completeness is defined as the number of detections correctly identified as clusters over the total number of mock clusters, as a function of redshift and intrinsic richness. Thus it provides a measure of how many objects are lost in the detection procedure. On the other hand, the purity is a measure of the contamination level of the cluster sample. It is defined as the fraction of detections that match with the clusters in the mock catalogue, over the total number of detections, in a given bin of redshift and intrinsic richness. As shown in Fig.\ \ref{selfunc}, it turns out that the catalogue is highly pure, with a purity approaching 100\% over the whole redshift range for $\lambda^*\geq20$. \\
\indent In order to account for the selection function in the cluster counts modelling, we build a new dataset by applying the purity and the completeness to the real cluster catalogue. This dataset will be used to derive the multiplicative weights that will be considered in the cluster counts model, as we will detail in the following. Since we define the purity as a function of the observed intrinsic richness, $\lambda^*_{\rm ob}$, we assign each object in the real catalogue to a bin of observed intrinsic richness, in which we computed the purity. Subsequently, we extract a uniform random number between 0 and 1, and if it is lower than the purity corresponding to the bin, the object is considered in the aforementioned new dataset. Otherwise, it is rejected. In this way, the final sample will statistically take into account the effects of impurities. On the other hand, since the completeness is defined in bins of true intrinsic richness, $\lambda^*_{\rm tr}$, it is required to implement a method that assigns a value of completeness to an observed value of intrinsic richness. For this purpose, we derive several probability distributions from the mock catalogue describing the probability to obtain a true value of $\lambda^*$, given a range of observed intrinsic richness defined by $\lambda^{*\,\text{low}}_{\rm ob}$ and $\lambda^{*\,\text{up}}_{\rm ob}$, namely $P(\lambda^*_{\rm tr}|\lambda^{*\,\text{low}}_{\rm ob},\lambda^{*\,\text{up}}_{\rm ob})$. 
We find that these distributions are reasonably Gaussian. Then, given a galaxy cluster in our dataset with a value of $\lambda^*_{\rm ob}$ in a given range, we perform a Gaussian Monte Carlo extraction from $P(\lambda^*_{\rm tr}|\lambda^{*\,\text{low}}_{\rm ob},\lambda^{*\,\text{\rm up}}_{\rm ob})$ through which we obtain a value of $\lambda^*_{\rm tr}$. Given the extracted true value of intrinsic richness, we assign a completeness value to the considered object.
Having this new catalogue corrected for the purity and the completeness, we construct a weight factor defined as the ratio between the uncorrected counts and the corrected ones, for bins in intrinsic richness (denoted by $\Delta\lambda^*_{\text{ob},i}$) and redshift (labelled as $\Delta z_{\text{ob},j}$) . These weight factors, $w(\Delta{\lambda^*_{\text{ob},i}},\Delta z_{\text{ob},j})$, will be used to weight the counts model as described in Section \ref{clustermodel}.
The value of $w(\Delta{\lambda^*_{\text{ob},i}},\Delta z_{\text{ob},j})$ in the first bins of intrinsic richness amounts to $\sim0.87$, $\sim0.76$, $\sim0.64$, in the redshift bins $z\in[0.10,\,0.30]$, $z\in[0.30,\,0.45]$, $z\in[0.45,\,0.60]$, respectively, while we derive no correction for the other bins (i.e.\ in these bins the weights are equal to 1). \\
\indent The measured counts of the AMICO KiDS-DR3 clusters are shown in Fig.\ \ref{figCounts}, along with the 68\% confidence level derived in Section \ref{SecResults}.

\section{Modelling}\label{SectionMethods}
\subsection{Model for the weak-lensing masses}\label{sectionLensingModelling}
\indent We model the scaling relation between the estimated cluster mean masses and the intrinsic richnesses using the following functional form:
\begin{align}\label{relscala}
\log\frac{M_{200}}{10^{14}M_\odot/h} &= \alpha + \beta \int_0^\infty {\rm d}\lambda^*\,\, P(\lambda^*|\lambda^*_{\rm eff})\,\log \frac{\lambda^*}{\lambda^*_{\rm piv}} + \nonumber\\ &+\gamma \int_0^\infty {\rm d}z\,\, P(z|z_{\rm eff})\,\log \frac{E(z)}{E(z_{\rm piv})}\,\,,
\end{align}
where $E(z)\equiv H(z)/H_0$, while $z_{\text{eff}}$ and $\lambda^*_{\text{eff}}$ are the lensing-weighted effective redshift and richness, respectively, whose computation is described in \citet{lensing}. The probability distributions $P(\lambda^*|\lambda^*_{\rm eff})$ and $P(z|z_{\rm eff})$ are assumed to be Gaussian, with mean equal to the values of $\lambda^*_{\rm eff}$ and $z_{\rm eff}$ listed in Table \ref{tab2}, and rms given by the uncertainties on $\lambda^*_{\rm eff}$ and $z_{\rm eff}$, respectively. The last term in Eq.\ \eqref{relscala} accounts for deviations in the redshift evolution from what is predicted in the self-similar growth scenario \citep{serenoettori}. Following \citet{lensing}, we set $\lambda^*_{\rm piv} = 30$ and $z_{\rm piv}=0.35$. In Eq.\ \eqref{relscala} the observables are the estimated mean mass values, $\log\bar{M}_{200}$, shown in Table \ref{tab2}, along with the effective values of redshift, $z_\text{\rm eff}$, and of intrinsic richness, $\lambda_\text{\rm eff}$, in the given bin. Furthermore, since $\log\bar{M}_{200}$ depends on cosmological parameters, we adopt the rescaling described in \citet{rescaling}, i.e.:
\begin{equation}\label{resc}
    \bar{M}_{200,\,\rm new}=\bar{M}_{200,\,\rm ref}\;\frac{\left[D_d^{-\frac{3\delta\gamma}{2-\delta\gamma}}\left(\frac{D_{ds}}{D_s}\right)^{-\frac{3}{2-\delta\gamma}}H(z)^{-\frac{1+\delta\gamma}{1-\delta\gamma/2}}\right]_{\rm new}}{\left[D_d^{-\frac{3\delta\gamma}{2-\delta\gamma}}\left(\frac{D_{ds}}{D_s}\right)^{-\frac{3}{2-\delta\gamma}}H(z)^{-\frac{1+\delta\gamma}{1-\delta\gamma/2}}\right]_{\rm ref}},
\end{equation}
where $ref$ indicates the assumed reference cosmology, i.e.\ $\Omega_{\rm m}=0.3$ and $h=0.7$, while the subscript $new$ refers to the test one. We set the slope $\delta\gamma=0$, corresponding to the case of a singular isothermal profile, this being a good approximation in general \citep[as discussed in][]{rescaling}. Given e.g.\ $M_{200}\simeq 10^{15}\,\,M_\odot$ and $c_{200}\simeq3$, $\delta\gamma\simeq-0.1$ is obtained, thus we varied $\delta\gamma$ in the reasonable range $[-0.2,0.2]$ and verified that this does not impact on the final results. The terms $D_{\rm s}$, $D_{\rm d}$, and $D_{\rm ds}$ are the source, the lens and the lens-source angular diameter distances, respectively. In $D_{\rm d}$, $D_{\rm ds}$ and in the Hubble parameter, $H(z)$, we assume the effective redshift values, $z_\text{eff}$, listed in Table \ref{tab2}. With regard to the redshifts of the sources, we consider the lensed source effective redshifts, $z_{\text{s,eff}}$, listed in Table \ref{tab2}, obtained by following the procedure described in \citet{giocoli2020}. In particular, we obtain $z_{\text{s,eff}}$ by weighting the redshift of each source by the corresponding source density for each derived radial bin for each cluster. We then consider the mean value of source redshift in bins of cluster richness per redshift. We verified that we can neglect the uncertainty on $z_{\text{s,eff}}$ in our analysis. In such mass rescaling, the relative uncertainty on $\log\bar{M}_{200,\,\rm new}$ is constant, corresponding to the relative errors on $\log\bar{M}_{200,\,\rm ref}$.

\subsection{Model for the cluster counts}\label{clustermodel}
\begin{table*}[t]
\caption{\label{tab1}Parameters considered in the joint analysis of cluster counts and stacked weak-lensing data.}
  \centering
    \begin{tabular}{l c c r} 
      Parameter & Description & Prior & Posterior \\ 
      \hline
      \rule{0pt}{4ex}
      $\Omega_{\rm m}$ & Total matter density parameter & [0.09, 1] & $\omegam^{+\omegamUp}_{-\omegamLow}$\\ \rule{0pt}{2.5ex}
      $\sigma_8$ & Amplitude of the matter power spectrum & [0.4, 1.5] & $\sigmaotto^{+\sigmaUp}_{-\sigmaLow}$\\ \rule{0pt}{2.5ex}
      $S_8\equiv\sigma_8(\Omega_m/0.3)^{0.5}$ & Cluster normalisation parameter & --- & $\Sotto^{+\SUp}_{-\SLow}$\\ \rule{0pt}{2.5ex}
      $\alpha$ & Normalisation of the mass-observable scaling relation & [-2, 2] & $\ascal^{+\aup}_{-\alow}$\\ \rule{0pt}{2.5ex}
      $\beta$ & Slope of the mass-observable scaling relation & [0, 5] & $\bscal^{+\bup}_{-\blow}$ \\ \rule{0pt}{2.5ex}
      $\gamma$ & Redshift evolution of the mass-observable scaling  relation & [-4, 4] & $\cscal^{+\cscalup}_{-\clow}$\\ \rule{0pt}{2.5ex}
      $\sigma_{\rm intr,0}$ & Normalisation of $\sigma_{\rm intr}$ & [0.05, 1] & $\scatter^{+\scatterUp}_{-\scatterLow}$\\ \rule{0pt}{2.5ex}
      $\sigma_{\rm intr,\lambda^*}$ & $\lambda^*$ evolution of $\sigma_{\rm intr}$ & [-5, 5] & $\scatterM^{+\scatterMUp}_{-\scatterMLow}$\\ \rule{0pt}{2.5ex}
      $\Omega_{\rm b}$ & Baryon density parameter & $\mathcal{N}(0.0486, 0.0017)$ & ---\\ \rule{0pt}{2.5ex}
      $\tau$ & Thomson scattering optical depth at reionization & $\mathcal{N}(0.0544, 0.0365)$ & ---\\ \rule{0pt}{2.5ex}
      $n_{\rm s}$ & Primordial power spectrum spectral index & $\mathcal{N}(0.9649, 0.0210)$ & --- \\ \rule{0pt}{2.5ex}
      $h\equiv H_0/(100$ km/s/Mpc) & Normalised Hubble constant & $\mathcal{N}(0.7, 0.1)$ & --- \\ \rule{0pt}{2.5ex}
      $s$ & Slope correction to the halo mass function & $\mathcal{N}(0.037,0.014)$ & --- \\ \rule{0pt}{2.5ex}
      $q$ & Amplitude correction to the halo mass function & $\mathcal{N}(1.008,0.019)$ & --- \\ \rule{0pt}{2.5ex}
      $\boldsymbol{\delta}_{\rm b}$ & Fluctuation of the matter density due to super-survey modes & Gaussian\tnote{*} & --- \\
    \end{tabular}
  \tablefoot{In the third column the priors on the parameters are listed, and in particular a range represents a uniform prior, while $\mathcal{N}(\mu,\sigma)$ stands for a Gaussian prior with mean $\mu$ and standard deviation $\sigma$. The Gaussian prior on $\boldsymbol{\delta}_{\rm b}$ is cosmology-dependent: both the mean and the standard deviation change at each MCMC step. In the fourth column we show the median values of the 1-D marginalised posteriors, along with the 16-th and 84-th percentiles. The posterior distributions of $\Omega_b$, $\tau$, $n_s$, $h$, $s$, $q$ and $\boldsymbol{\delta}_b$, are not shown since these nuisance parameters are not constrained in our analysis.}
\end{table*}
The specific characteristics of the dataset must be included in the model and in the covariance matrix of the likelihood function. We describe the expectation value of the counts in a given bin of intrinsic richness, $\Delta\lambda^{*}_{\text{ob},i}$, and of observed redshift, $\Delta z_{\text{ob},j}$, as
\begin{align}\label{themodel}
\langle N(\Delta{\lambda^*_{\text{ob},i}},&\Delta z_{\text{ob},j})\rangle = w(\Delta{\lambda^*_{\text{ob},i}},\Delta z_{\text{ob},j})\,\,\Omega \int\displaylimits_{0}^{\infty} \dd z_{\rm tr}\,\,\frac{\dd V}{\dd z_{\rm tr}\dd\Omega}\times\nonumber\\
&\times\int\displaylimits_{0}^{\infty} \dd M_{\rm tr} \,\,\frac{\dd n(M_{\rm tr},z_{\rm tr})}{\dd M_{\rm tr}}\,\, \int\displaylimits_{0}^{\infty}\dd\lambda^*_{\rm tr}\,\,P(\lambda^*_{\rm tr}| M_{\rm tr},z_{\rm tr})\,\times\nonumber\\ &\times\int\displaylimits_{\Delta z_{\text{ob},j}}\dd z_{\rm ob} \,\,P(z_{\rm ob}|z_{\rm tr,\,corr})\,\int\displaylimits_{\Delta\lambda^*_{\text{ob},i}}\dd \lambda^*_{\rm ob} \,\,P(\lambda^*_{\rm ob}|\lambda^*_{\rm tr})\,,
\end{align}
where $z_{\rm tr}$ is the true redshift, $V$ is the comoving volume, $\Omega$ is the survey effective area, $M_{\rm tr}$ the true mass, and $\dd n(M_{\rm tr},z_{\rm tr})/\dd M_{\rm tr}$ is the mass function, for which the model by \citet{tinker} is assumed.  The term $w(\Delta{\lambda^*_{\text{ob},i}},\Delta z_{\text{ob},j})$ is the weight factor described in Section \ref{SectionMock}, accounting for the purity and completeness of the sample. The probability distribution $P(z_{\rm ob}|z_{\rm tr,\,corr})$, assessed through the mock catalogue described in Section \ref{SectionMock}, is a Gaussian accounting for the uncertainties on the redshifts. The mean of such distribution, $z_{\rm tr,\,corr}$, is the true redshift corrected by the redshift bias, and it is expressed as
\begin{equation}
z_{\rm tr,\,corr} = z_{\rm tr} + \Delta z_{\rm bias}\,(1+z_{\rm tr})\,,
\end{equation}
where $\Delta z_{\rm bias}\,(1+z_{\rm tr})$ is the redshift bias term discussed in \citet{catalogue}, with $\Delta z_{\rm bias}=0.02$. In particular, this bias corresponds to what was found in \citet{dejong} by comparing the KiDS photo-$z$s to the GAMA spectroscopic redshifts (see their Table 8). In order to assess the impact of its uncertainty, we included $\Delta z_{\rm bias}$ as a free parameter in the model, assuming a Gaussian prior with mean equal to $0.02$ and rms equal to $0.02$, which is similar to the rms of the sample and much larger than the rms of the mean \citep[see Fig.\ 7 in][]{catalogue}. As we verified, 
such uncertainty on $\Delta z_{\rm bias}$ does not significantly impact our final results. Conversely, AMICO provides unbiased estimates of redshift \citep[see][]{catalogue}, thus we model $P(z_{\rm ob}|z_{\rm tr})$ by keeping the mean of such distributions fixed to the central value of $\Delta z_{\rm tr}$. In particular, in the mock catalogue we measure $P(z_{\rm ob}|z_{\rm tr})$ in several bins of $z_{\rm tr}$, $\Delta z_{\rm tr}$, and perform the statistical MCMC analysis assuming a common flat prior on the rms in all the $\Delta z_{\rm tr}$ bins. The resulting rms of $P(z_{\rm ob}|z_{\rm tr})$ is equal to 0.025. AMICO also provides unbiased estimates for $\lambda^*$, thus following the same procedure adopted for $P(z_{\rm ob}|z_{\rm tr})$ we derive an uncertainty of $\sim17\%$ on $\lambda^*_{\rm ob}$, defining the rms of the Gaussian distribution $P(\lambda^*_{\rm ob}|\lambda^*_{\rm tr})$, whose mean is equal to $\lambda^*_{\rm tr}$. We neglect the uncertainties on the rms of $P(z_{\rm ob}|z_{\rm tr})$ and $P(\lambda^*_{\rm ob}|\lambda^*_{\rm tr})$, amounting to $\sim1\%$, since we verified their negligible effect on the final results. \\
\indent Furthermore, $P(\lambda^*_{\rm tr}|M_{\rm tr},z_{\rm tr})$ is a probability distribution that weights the expected counts according to the shape of the mass-observable scaling relation, and it is expressed as follows:
\begin{equation}\label{tre}
P(\lambda^*_{\rm tr}|M_{\rm tr},z_{\rm tr})= \frac{P(M_{\rm tr}|\lambda^*_{\rm tr},z_{\rm tr})\,P(\lambda^*_{\rm tr}|z_{\rm tr})}{P( M_{\rm tr}|z_{\rm tr})},
\end{equation}
where the distribution $P(M_{\rm tr}|\lambda^*_{\rm tr},z_{\rm tr})$ is a log-normal whose mean is given by the mass-observable scaling relation and the standard deviation is given by the \textit{intrinsic scatter}, $\sigma_{\rm intr}$, set as a free parameter of the model: 
\begin{equation}
P(\log M_{\rm tr}|\lambda^*_{\rm tr},z_{\rm tr})=\frac{1}{\sqrt{2\pi}\sigma_{\rm intr}}\exp\left(-\frac{x^2(M_{\rm tr},\lambda^*_{\rm tr},z_{\rm tr})}{2\sigma^2_{\rm intr}}\right),
\end{equation}
where 
\begin{align}
x(M_{\rm tr},\lambda^*_{\rm tr},z_{\rm tr})=&\log\frac{M_{\rm tr}}{10^{14}M_\odot/h}\,\,- \nonumber\\
&-\,\Bigg(\alpha+\beta\log\frac{\lambda_{\rm tr}^*}{\lambda^*_{\rm piv}}+\gamma\log\frac{E(z_{\rm tr})}{E(z_{\rm piv})}\Bigg),\,
\end{align}
and
\begin{equation}
\sigma_{\rm intr} = \sigma_{\rm intr,0}+\sigma_{\rm intr,\lambda^*}\log\frac{\lambda^*_{\rm tr}}{\lambda^*_{\rm piv}}.
\end{equation}
The distribution $P(M_{\rm tr}|\lambda^*_{\rm tr},z_{\rm tr})$, indeed, accounts for the intrinsic uncertainty that affects a scaling relation between the intrinsic richness and the mass: given an infinitely accurate scaling relation, represented by the mean, the cluster mass provided by a value of intrinsic richness is scattered from the true value. 
Furthermore, $P(\lambda^*_{\rm tr}|\Delta z_{\rm tr})$ in Eq.\ \eqref{tre} is a power-law with an exponential cut-off, derived from the mock catalogue by considering the objects with $\lambda^*_{\rm tr}\gtrsim 20$. Specifically, similarly to other literature analyses \citep[e.g.][]{murata19,costanzi,abbott20}, $P(\lambda^*_{\rm tr}|M_{\rm tr},z_{\rm tr})$ is assumed to be cosmology-independent. Thus we assume that the ratio $P(\lambda^*_{\rm tr}|z_{\rm tr})/P(M_{\rm tr}|z_{\rm tr})$ is cosmology-independent, where $P(M_{\rm tr}|z_{\rm tr})$ acts as a normalisation of $P(\lambda^*_{\rm tr}|M_{\rm tr},z_{\rm tr})$:
\begin{equation}
P(M_{\rm tr}|z_{\rm tr})=\int\displaylimits_{0}^{\infty}
\dd \lambda^*_{\rm tr}\, P(M_{\rm tr}|\lambda^*_{\rm tr},z_{\rm tr})\,P(\lambda^*_{\rm tr}|z_{\rm tr})\,.
\end{equation}

\subsection{Halo mass function systematics}\label{MFuncertainty}
As mentioned in Section \ref{clustermodel}, we assume the \citet{tinker} halo mass function to model the observed cluster counts. Following \citet{costanzi}, in order to characterize the systematic uncertainty in the halo mass function in dark matter only simulations, we relate the \citet{tinker} mass function to the true mass function via
\begin{equation}
n(M,z)=n(M,z)^{\rm Tinker}(s\log (M/M^*)+q),
\end{equation}
where $\log M^*=13.8$ $h^{-1}$M$_\odot$ is the pivot mass, while $q$ and $s$ are free parameters of the model with Gaussian prior having the following covariance matrix
\begin{equation}\label{covMtinker}
C(\bar{s},\bar{q}) = 
\begin{pmatrix}
0.00019 & 0.00024 \\
0.00024 & 0.00038
\end{pmatrix}
,
\end{equation}
and $\bar{s}=0.037$, $\bar{q}=1.008$ as the mean values. Diagonalising the matrix \eqref{covMtinker} we obtain the following 1-D Gaussian priors: $\mathcal{N}(\bar{s},\sigma_{\rm s})=\mathcal{N}(0.037,0.014)$, $\mathcal{N}(\bar{q},\sigma_{\rm q})=\mathcal{N}(1.008,0.019)$, where $\mathcal{N}(\mu,\sigma)$ stands for a Gaussian distribution with mean $\mu$ and standard deviation $\sigma$.

\subsection{Likelihood}\label{seclikelihood}
\begin{figure*}[t]
   \centering
   \includegraphics[width=18cm,height=18cm]{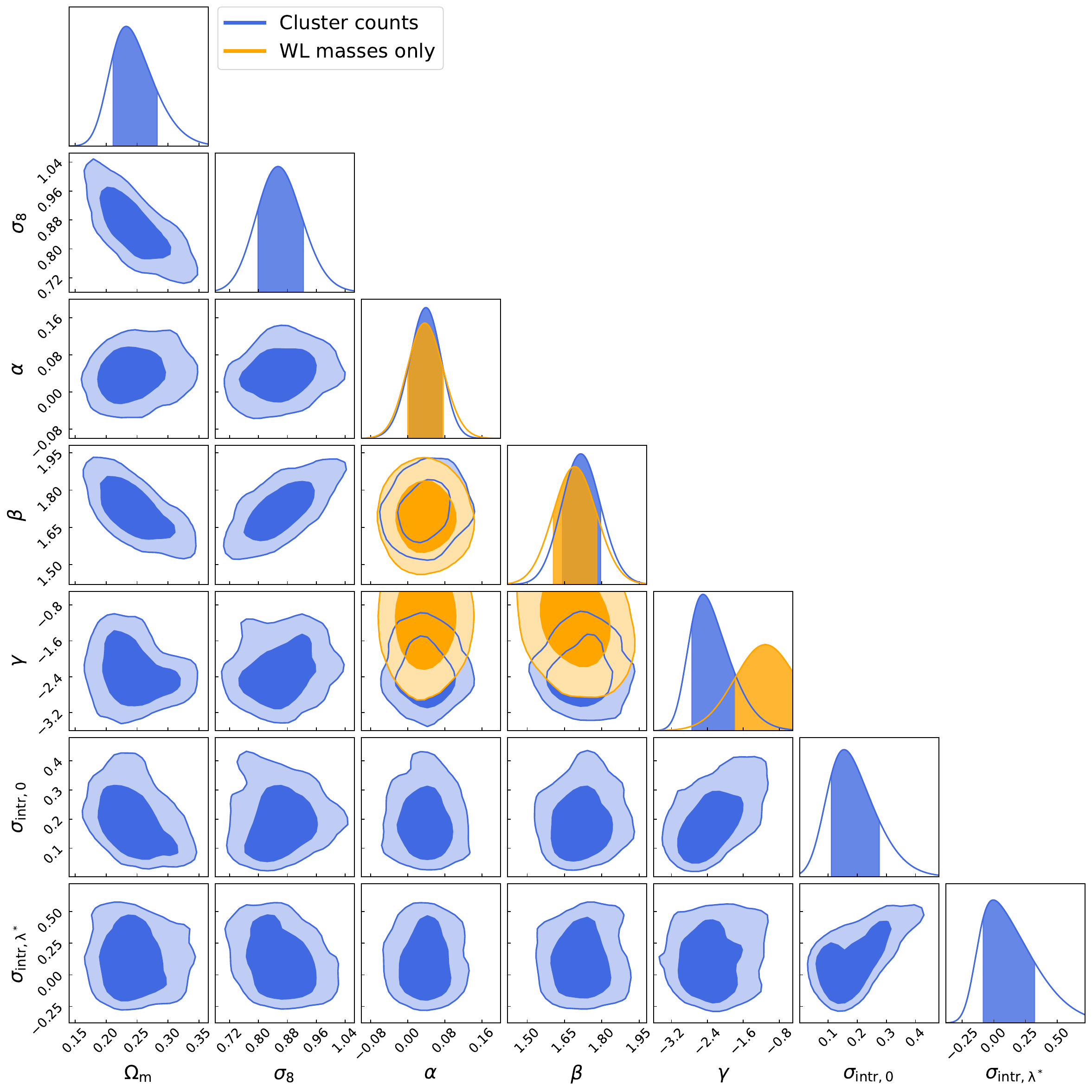}
      \caption{Constraints on $\Omega_{\rm m}$, $\sigma_8$, $\alpha$, $\beta$, $\gamma$, $\sigma_{\rm intr,0}$, $\sigma_{\rm intr,\lambda^*}$, derived in a flat $\Lambda$CDM universe by combining the redshift bins $z\in [0.1,\,0.3]$, $z\in[0.3,\,0.45]$, $z\in[0.45,\,0.6]$ and assuming a minimum intrinsic richness $\lambda^*_{\rm ob,\,min}=20$ for cluster counts. The shown posteriors are marginalised also over $\Omega_{\rm b}$, $\tau$, $n_{\rm s}$, $h$, $s$, $q$, and $\boldsymbol{\delta}_{\rm b}$. The blue contours represent the results obtained from the joint analysis of cluster counts and weak-lensing masses, while the orange contours show the posteriors on $\alpha$, $\beta$ and $\gamma$, derived from the analysis including only the weak-lensing masses as in \citet{lensing}. The confidence ellipses correspond to 68\% and 95\%, while the bands over the 1-D marginalised posteriors represent the 68\% of confidence.}
         \label{fig2}
   \end{figure*}
Our likelihood function encapsulates the description of cluster counts and weak-lensing masses. We base the likelihood term describing the counts, $\Lagr_{\rm counts}$, on the functional form given by \citet{lacasa}, that is a convolution of a Poissonian likelihood describing the counts, and a Gaussian distribution accounting for the super-sample covariance (SSC):
\begin{align}\label{eqLikelihood}
\Lagr_{\rm counts}=&\int \dd\boldsymbol{\delta}_{\rm b}^{n_{\rm z}}
\left[\prod_{i,j}\text{Poiss}\left(N_{i,j}| \bar{N}_{i,j} +
  \frac{\partial N_{i,j}}{\partial\delta_{\text{b},j}}\delta_{\text{b},j} \right)
  \right]\,\, \mathcal{N}(\boldsymbol{\delta}_{\rm b}|0,S)\,.
\end{align}
In the equation above, $\mathcal{N}(\boldsymbol{\delta}_{\rm b}|0,S)$ is the Gaussian function describing the SSC effects on cluster counts measurements, which is a function of the matter density contrast fluctuation, $\boldsymbol{\delta}_{\rm b}$, and has covariance matrix $S$\footnote{For the computation of the $S$ matrix, we refer to the codes at \url{https://github.com/fabienlacasa/PySSC} developed by \citet{lacasa}}. In particular, $n_{\rm z}$ is the number of redshift bins considered in the modelling procedure, and it defines the dimension of the integration variable, $\boldsymbol{\delta}_{\rm b}=\{\delta_{\text{b},1},...,\delta_{\text{b},n_{\rm z}}\}$, and of the $S$ matrix, whose dimension is $n_{\rm z}\times n_{\rm z}$. Thus each $\delta_{\text{b},j}$ represents the fluctuation of the measured matter density contrast, with respect to the expected one, in a given bin of redshift. The indices $i$ and $j$ are the labels of the bins of intrinsic richness and redshift, while $N_{i,j}\equiv
N(\Delta\lambda^*_{\text{ob},i},\Delta z_{\text{ob},j})$ is the observed cluster number counts in a bin of intrinsic richness and redshift, and $\bar{N}_{i,j}$ is the model defined in Eq.\ \eqref{themodel}. The term $\partial N_{i,j}/\partial\delta_{\text{b},j}$ is the response of the counts, i.e.\ the measure of how the counts vary with changes of the background density, and it is expressed as:
\begin{align}
\frac{\partial N_{i,j}}{\partial\delta_{\text{b},j}}& = w(\Delta{\lambda^*_{\text{ob},i}},\Delta z_{\text{ob},j})\,\,\Omega \int\displaylimits_{0}^{\infty} \dd z_{\rm tr}\,\,\frac{\dd V}{\dd z_{\rm tr}\dd\Omega}\times\nonumber\\
&\times\int\displaylimits_{0}^{\infty} \dd M_{\rm tr} \,\,\frac{\dd n(M_{\rm tr},z_{\rm tr})}{\dd M_{\rm tr}}\,\,b(M_{\rm tr},z_{\rm tr})\,\, \int\displaylimits_{0}^{\infty}\dd\lambda^*_{\rm tr}\,\,P(\lambda^*_{\rm tr}| M_{\rm tr},z_{\rm tr})\,\times\nonumber\\ &\times\int\displaylimits_{\Delta z_{\text{ob},j}}\dd z_{\rm ob} \,\,P(z_{\rm ob}|z_{\rm tr,\,corr})\,\int\displaylimits_{\Delta\lambda^*_{\text{ob},i}}\dd \lambda^*_{\rm ob} \,\,P(\lambda^*_{\rm ob}|\lambda^*_{\rm tr})\,,
\end{align}
that is, the response is similar to the model described in
Eq.\ \eqref{themodel}, in which we also include the contribution of
the linear bias $b(M,z)$. \\
\indent For computational purposes, we consider in the analysis an alternative form of the likelihood, $\Lagr'_{\rm counts}$, that is the integrand in Eq.\ \eqref{eqLikelihood}, of which we compute the natural logarithm:
\begin{equation}\label{eqLikelihood2}
\ln\Lagr'_{\rm counts}=\ln\left[\prod_{i,j}\text{Poiss}\left(N_{i,j}| \bar{N}_{i,j}
  + \frac{\partial N_{i,j}}{\partial\delta_{\text{b},j}}\delta_{\text{b},j} \right)
  \cdot \mathcal{N}(\boldsymbol{\delta}_\text{b}|0,S)\right].
\end{equation}
\begin{figure*}[t!]
\centering
\includegraphics[width = 0.4 \hsize, height = 7cm] {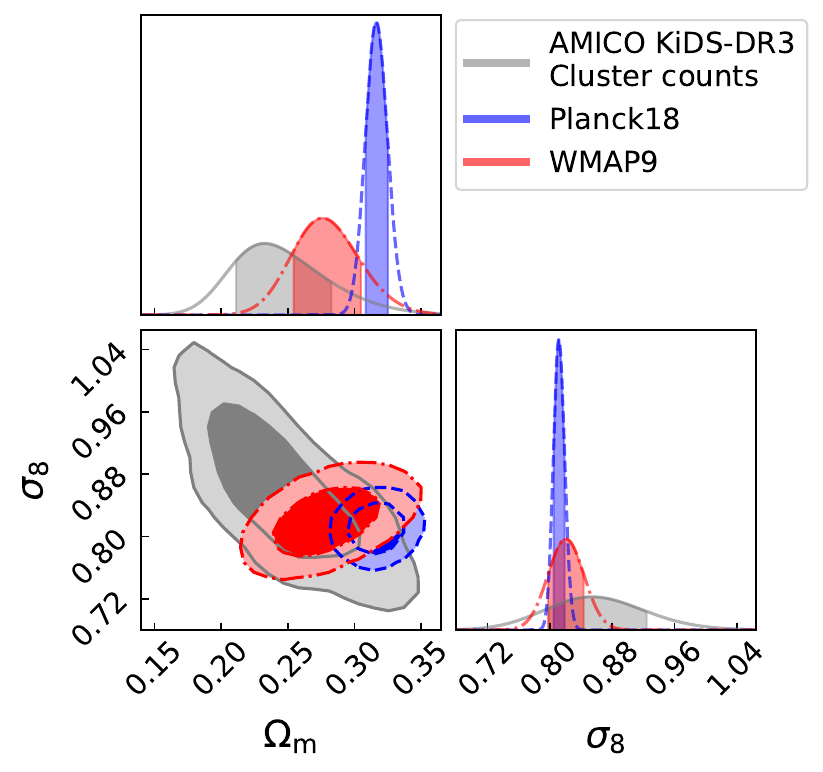}
\includegraphics[width = 0.4 \hsize, height = 7cm] {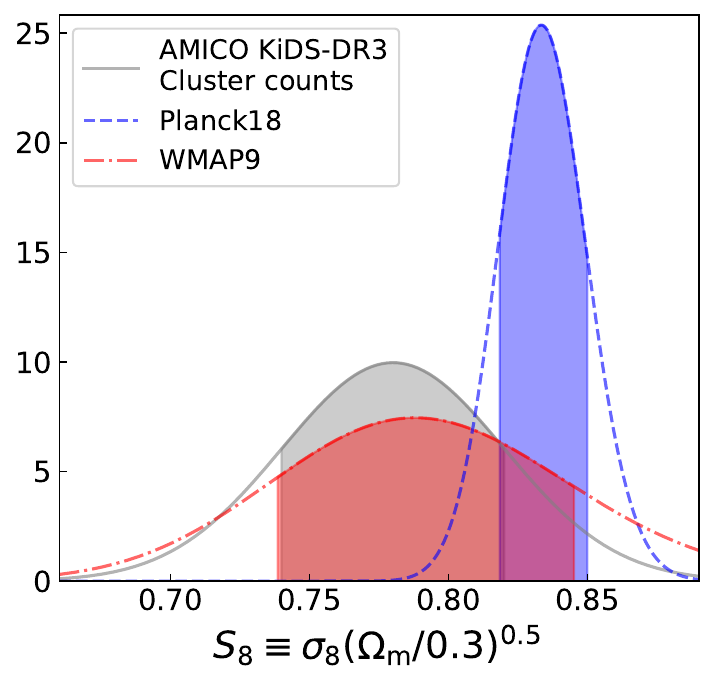}
\caption{Comparison with WMAP and Planck results. In the left panel we show the $\Omega_{\rm m}$-$\sigma_8$ parameter space, along with the 1-D marginalised posteriors with the relative intervals between 16-th and 84-th percentiles, in the case of the cluster counts analysis in the AMICO KiDS-DR3 catalogue (grey solid lines). We also display, in the same panel, the results from WMAP \citep{wmap} (Table 3, WMAP-only Nine-year; red dash-dotted lines), and Planck \citep{planck} (Table 2, TT,TE,EE+lowE; blue dashed lines). In the right panel we show the posteriors for the parameter $S_8$, where the bands show the intervals between 16-th and 84-th percentiles. The symbols are the same as in the left panel.}
\label{common}
\end{figure*}
Here, we set
$\boldsymbol{\delta}_\text{b}=\{\delta_{\text{b},1},...,\delta_{\text{b},n_\text{z}}\}$ as free parameters of the model, with a multivariate Gaussian prior having $S$ as the covariance matrix. Due to the dependence on cosmological parameters of the $S$ matrix, the values of its elements change at every step of the MCMC. In turn, a variation of $S$ implies the change of the prior on $\boldsymbol{\delta}_\text{b}$. At the end of the MCMC, we marginalise over $\boldsymbol{\delta}_\text{b}$ to derive the posteriors of our parameters of interest. \\
\indent With regard to the likelihood describing the weak-lensing masses, $\Lagr_{\rm lens}$, we assume a log-normal functional form and then we consider its natural logarithm:
\begin{equation}\label{lagr2}
\ln\Lagr_{\rm lens}\propto\sum_{k=1}^{N_{\rm bin}}\sum_{l=1}^{N_{\rm bin}}[\log\bar{M}_{\rm ob}^{k}-\log\bar{M}_{\rm mod}^{l}]\,\boldsymbol{C}^{-1}_{M,kl}\,[\log\bar{M}_{\rm ob}^{l}-\log\bar{M}_{\rm mod}^{k}]\,,
\end{equation}
where $N_{\rm bin}$ corresponds to the number of bins in which the mean mass $\bar{M}_{200}$, or $\bar{M}_{\rm ob}$, is measured, through the weak-lensing analysis described in Section \ref{SectionLensing}. Furthermore, $\bar{M}_{\rm mod}$ represents the mass obtained from the scaling relation model described in Eq.\ \eqref{relscala}, where we assume the effective redshift and intrinsic richness values $z_{\rm eff}$ and $\lambda^*_{\rm eff}$ listed in Table \ref{tab2}. The covariance matrix $\boldsymbol{C}_M$ in Eq.\ \eqref{lagr2} has the following form:
\begin{equation}\label{covmatrixlensing}
\boldsymbol{C}_{M,kl}=\delta_{kl}E^2_k+[\sigma_{\rm sys}/\ln(10)]^2+\delta_{kl}(\sigma_{\rm intr}/\sqrt{N_{\rm cl}})^2\,,
\end{equation}
where $E_k$ represents the statistical error on $\log\bar{M}_{\rm ob}$ derived from the posterior distribution of $\log\bar{M}_{\rm ob}$, where we stress that the relative uncertainties are constants after the rescaling described in Section \ref{SectionLensing}. The term $\sigma_{\rm sys}=0.076$ is the sum in quadrature of the uncertainties on background selection, photo-$z$s, shear measurements, halo model, orientation and projections, obtained in \citet{lensing}, and $N_{\rm cl}$ is the number of clusters in the bin of intrinsic richness and redshift in which the mean mass has been derived. By dividing $\sigma_{\rm intr}$ by $\sqrt{N_{\rm cl}}$, we neglect the cluster clustering contribution to the last term of Eq.\ \eqref{covmatrixlensing}. \\
\indent Thus the logarithm of the joint likelihood, $\ln\Lagr$, is given by
\begin{equation}
\ln\Lagr=\ln\Lagr'_{\rm counts}+\ln\Lagr_{\rm lens}\,.
\end{equation}

\section{Results}\label{SecResults}
\begin{figure}[t!]
\centering\includegraphics[width = \hsize-0.7cm, height = 11.5cm] {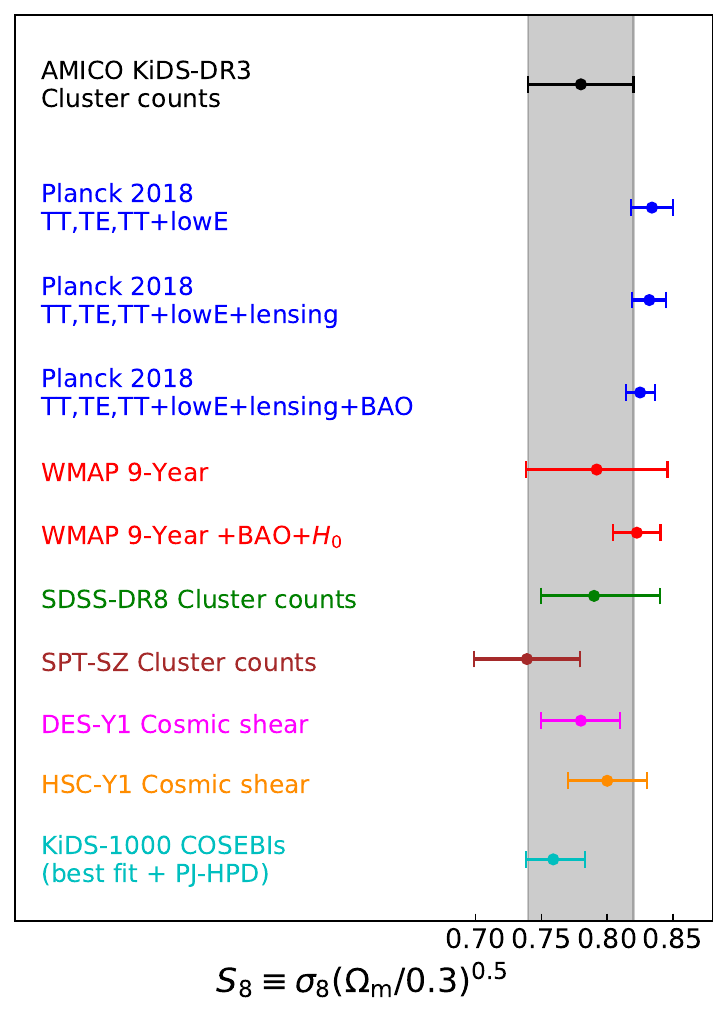}
\caption{Comparison of the constraints on $S_8\equiv\sigma_8(\Omega_\text{m}/0.3)^{0.5}$ obtained, from top to bottom, from the joint analysis of cluster counts and weak-lensing in the AMICO KiDS-DR3 catalogue (black dot), from the results obtained by \citet{planck} (blue dots), \citet{wmap} (red dots), \citet{costanzi} (green dot), \citet{bocquet19} (brown dot), \citet{troxel}  (magenta dot), \citet{hsc} (orange dot), \citet{asgari2020} (cyan dot). Median, 16-th and 84-th percentiles are shown.}
\label{external_datasets}
\end{figure}

We perform a cosmological analysis of cluster number counts and stacked weak-lensing based on the assumption of a flat $\Lambda$CDM model. The aim is to constrain the matter density parameter, $\Omega_{\rm m}$, and the square root of the mass variance computed on a scale of 8 Mpc$/h$, $\sigma_8$, both provided at $z=0$, along with the parameters defining the scaling relation between masses and intrinsic richnesses, $\alpha$, $\beta$, $\gamma$ in Eq.\ \eqref{relscala}, and the intrinsic scatter, $\sigma_{\rm intr}$. Therefore we set $\Omega_\text{m}$, $\sigma_8$, $\alpha$, $\beta$, $\gamma$, $\sigma_{\rm intr}$, as free parameters of the model, Eq.\ \eqref{themodel}, as well as the baryon density, $\Omega_\text{b}$, the optical depth at reionization, $\tau$, the primordial spectral index, $n_\text{s}$, the normalised Hubble constant, $h$, the Tinker mass function correction parameters, $q$ and $s$, described in Section \ref{MFuncertainty}, and the fluctuation of the mean density of matter due to super-survey modes, $\boldsymbol{\delta}_\text{b}$. We assume flat priors for $\Omega_\text{m}$, $\sigma_8$, $\alpha$, $\beta$, $\gamma$, $\sigma_{\rm intr}$, while we set Gaussian priors on the other parameters (see Table \ref{tab1}). In particular, for the Gaussian prior distributions of $\Omega_\text{b}$, $\tau$, $n_\text{s}$, $h$, we refer to the values obtained by \citet{planck} (Table 2, TT,TE,EE+lowE+lensing), assuming the same mean values and imposing a standard deviation equal to 5$\sigma$ for all the aforementioned parameters but $h$, for which we assume a standard deviation equal to 0.1. In this baseline cosmological model, we also assume three neutrino species, approximated as two massless states and a single massive neutrino of mass $m_\nu=0.06$ eV, following \citet{planck}. Finally, we assume a multivariate Gaussian prior for $\boldsymbol{\delta}_\text{b}$, as described in Section \ref{seclikelihood}.\\
\indent In our analysis we constrain the value of the \textit{cluster normalisation parameter}, $S_8 \equiv \sigma_8(\Omega_\text{m}/0.3)^{0.5}$. The significance of this parameter is rooted in the degeneracy between $\sigma_8$ and $\Omega_m$, being defined along the $\sigma_8-\Omega_\text{m}$ confidence regions. Since the number of massive clusters increases with both $\sigma_8$ and $\Omega_\text{m}$, in order to hold the cluster abundance fixed at its observed value, any increase in $\sigma_8$ must be compensated by a decrease in $\Omega_\text{m}$, implying that $S_8$ is held fixed. \\
\indent From this modelling we obtain $\Omega_\text{m}=\omegam^{+\omegamUp}_{-\omegamLow}$, $\sigma_8=\sigmaotto^{+\sigmaUp}_{-\sigmaLow}$, $S_8=\Sotto^{+\SUp}_{-\SLow}$, where we quote the median, 16-th and 84-th percentiles, as shown in Fig.\ \ref{fig2} and Table \ref{tab1}. In Fig.\ \ref{fig2} we also show that the results on the mass-observable scaling relation retrieved from this analysis, i.e.\ for $\alpha$, $\beta$, $\gamma$, are in agreement within 1$\sigma$ with those obtained by performing the modelling of the weak-lensing data only, as carried out by \citet{lensing}. In particular, the inclusion of the cluster counts in the analysis provides tighter constraints on the slope $\beta$, governing also the slope of the cluster model at low values of $\lambda^*$. Additionally, the cluster counts redshift evolution brings to a more accurate estimate of $\gamma$. Lastly, we find a tight constraint on the intrinsic scatter, deriving $\sigma_{\rm intr,0}=\scatter^{+\scatterUp}_{-\scatterLow}$ and $\sigma_{\rm intr,\lambda^*}=\scatterM^{+\scatterMUp}_{-\scatterMLow}$, which confirms the reliability of $\lambda^*$ as a mass proxy. This result on $\sigma_{\rm intr}$ is consistent within $1\sigma$ with that derived in \citet{sereno2020} from a weak-lensing analysis of the sample of AMICO clusters in KiDS-DR3. \\
\indent As shown in Fig.\ \ref{common}, the constraints obtained for $S_8$ and $\sigma_8$ are in agreement within 1$\sigma$ with WMAP results \citep{wmap} (Table 3, WMAP-only Nine-year), and with Planck results \citep{planck} (Table 2, TT,TE,EE+lowE). With regard to $\Omega_{\rm m}$, we find an agreement within 1$\sigma$ with WMAP and a 2$\sigma$ tension with Planck. Furthermore, in Fig.\ \ref{external_datasets} we show the comparison with the $S_8$ constraints obtained from additional external datasets. In particular, we find an agreement within 1$\sigma$ with the results obtained from the cluster counts analyses performed by \citet{costanzi}, based on SDSS-DR8 data, and by \citet{bocquet19}, based on the 2500 deg$^2$ SPT-SZ Survey data, as well as with the results derived from the cosmic shear analyses performed by \citet{troxel} on DES-Y1 data, \citet{hsc} on HSC-Y1 data, and \citet{asgari2020} on KiDS-DR4 data. The constraint on $S_8$ obtained from the cluster counts and weak-lensing joint analysis in DES \citep{abbott20}, $S_8=0.65^{+0.04}_{-0.04}$, not shown in Fig.\ \ref{external_datasets}, is not consistent with our result. \\
\indent Following a more conservative approach, we repeat the analysis assuming the threshold in intrinsic richness $\lambda^*\geq20$ also for the weak-lensing data. This leads to $\Omega_{\rm m}=0.27^{+0.04}_{-0.05}$, $\sigma_8=0.83^{+0.06}_{-0.07}$, $S_8=0.78^{+0.04}_{-0.04}$, which are consistent within 1$\sigma$ with the constraints derived from the analysis previously described. Also for the other free parameters of the model the consistency within 1$\sigma$ still holds.

\section{Conclusions}\label{SecConclusions}
We performed a galaxy cluster abundance analysis in the AMICO KiDS-DR3 catalogue \citep{catalogue}, constraining simultaneously the cosmological parameters and the cluster mass-observable scaling relation. In particular, we relied on the intrinsic richness, defined in Eq.\ \eqref{lambda}, as the observable linked to the cluster masses. The sample exploited for cluster counts includes 3652 galaxy clusters having intrinsic richness $\lambda^*\geq20$, in the redshift bins $z\in[0.1,0.3]$, $z\in[0.3,0.45]$, $z\in[0.45,0.6]$. For the weak-lensing analysis we followed the procedure developed by \citet{lensing}, not assuming any threshold in $\lambda^*$. We assessed the incompleteness and the impurities of the cluster sample by exploiting a mock catalogue developed by \citet{catalogue}, and we corrected our data accordingly. \\
\indent We assumed a model for cluster counts, shown in Eq.\ \eqref{themodel}, accounting for the redshift uncertainties and for the mass-observable scaling relation. In particular, the mass-observable scaling relation plays a crucial role in the $P(\lambda^*_{\rm tr}|M_{\rm tr},z_{\rm tr})$ term given by Eq.\ \eqref{tre}, which also depends on the observed distribution of galaxy clusters as a function of the intrinsic richness. Furthermore, this term includes the contribution of the intrinsic scatter of the scaling relation, $\sigma_{\rm intr}$, which is considered as an unknown parameter. Subsequently, we modelled the cluster counts and the scaling relation by combining the relative likelihood functions. \\
\indent Assuming a flat $\Lambda$CDM model with massive neutrinos, we found $\Omega_\text{m}=\omegam^{+\omegamUp}_{-\omegamLow}$, $\sigma_8=\sigmaotto^{+\sigmaUp}_{-\sigmaLow}$, $S_8=\Sotto^{+\SUp}_{-\SLow}$, which are competitive constraints, in terms of uncertainties, with results of state-of-the-art cluster number count analyses. In addition, the result on $S_8$ is in agreement within 1$\sigma$ with the results from WMAP and Planck. We also derived results for the scaling relation that are consistent within 1$\sigma$ with those obtained by modelling only the weak-lensing signal as in \citet{lensing}, thus validating the reliability of our model. With regard to the intrinsic scatter, we found $\sigma_{\rm intr,0}=\scatter^{+\scatterUp}_{-\scatterLow}$ and $\sigma_{\rm intr,\lambda^*}=\scatterM^{+\scatterMUp}_{-\scatterMLow}$, which is a very competitive result compared to the present-day estimates in the field of galaxy clusters, outlining the goodness of the assumption of $\lambda^*$ as the mass proxy.\\
\indent In \citet{nanni2020} we analyse the AMICO KiDS-DR3 cluster clustering, to derive constraints on $S_8$ and on the mass-observable scaling relation. As the next step, we will combine counts, clustering and weak-lensing to improve further the accuracy of our results. Furthermore, we expect more accurate constraints on $S_8$ and on the mass-observable scaling relation from the analysis of the latest KiDS Data Release \citep[DR4,][]{Kuijken2019}. It covers an area of 1000 square degrees (more than two thirds of the final area), and photometry extends to the near infrared (ugriZYJHK$_s$) joining the data from KiDS and VIKING \citep{edge} surveys, thus allowing, for instance, to improve the photometric redshift estimates \citep{wright19}. 

\section*{Acknowledgements}
Based on data products from observations made with ESO Telescopes
at the La Silla Paranal Observatory under programme IDs 177.A-3016, 177.A3017 and 177.A-3018, and on data products produced by Target/OmegaCEN, INAF-OACN, INAF-OAPD and the KiDS production team, on behalf of the KiDS consortium. \\
\indent The authors acknowledge the use of computational resources from the parallel computing cluster of the Open Physics Hub (\url{https://site.unibo.it/openphysicshub/en}) at the Physics and Astronomy Department in Bologna. FM, LM and CG acknowledge the support from the grant ASI n.2018-23-HH.0. LM and CG also acknowledge the support from the grant PRIN-MIUR 2017 WSCC32. MS acknowledges financial contribution from contract ASI-INAF n.2017-14-H.0 and INAF mainstream project 1.05.01.86.10.  \\
\indent We thank Andrea Biviano, Matteo Costanzi, Catherine Heymans and Konrad Kuijken for their valuable advice. We also thank Fabien Lacasa for the support in the development of an optimal numerical implementation of the likelihood function describing the cluster counts, i.e.\ Eq.\ \ref{eqLikelihood2}.

\section*{Data availability}
The data underlying this article will be shared on reasonable request to the corresponding author.

\bibliography{sample}

\begin{thebibliography}{72}
\expandafter\ifx\csname natexlab\endcsname\relax\def\natexlab#1{#1}\fi

\bibitem[{{Abbott} {et~al.}(2020){Abbott}, {Aguena}, {Alarcon}, {Allam},
  {Allen}, {Annis}, {Avila}, {Bacon}, {Bechtol}, {Bermeo}, {Bernstein},
  {Bertin}, {Bhargava}, {Bocquet}, {Brooks}, {Brout}, {Buckley-Geer}, {Burke},
  {Carnero Rosell}, {Carrasco Kind}, {Carretero}, {Castander}, {Cawthon},
  {Chang}, {Chen}, {Choi}, {Costanzi}, {Crocce}, {da Costa}, {Davis}, {De
  Vicente}, {DeRose}, {Desai}, {Diehl}, {Dietrich}, {Dodelson}, {Doel},
  {Drlica-Wagner}, {Eckert}, {Eifler}, {Elvin-Poole}, {Estrada}, {Everett},
  {Evrard}, {Farahi}, {Ferrero}, {Flaugher}, {Fosalba}, {Frieman},
  {Garc{\'\i}a-Bellido}, {Gatti}, {Gaztanaga}, {Gerdes}, {Giannantonio},
  {Giles}, {Grandis}, {Gruen}, {Gruendl}, {Gschwend}, {Gutierrez}, {Hartley},
  {Hinton}, {Hollowood}, {Honscheid}, {Hoyle}, {Huterer}, {James}, {Jarvis},
  {Jeltema}, {Johnson}, {Johnson}, {Kent}, {Krause}, {Kron}, {Kuehn},
  {Kuropatkin}, {Lahav}, {Li}, {Lidman}, {Lima}, {Lin}, {MacCrann}, {Maia},
  {Mantz}, {Marshall}, {Martini}, {Mayers}, {Melchior}, {Mena-Fern{\'a}ndez},
  {Menanteau}, {Miquel}, {Mohr}, {Nichol}, {Nord}, {Ogando}, {Palmese},
  {Paz-Chinch{\'o}n}, {Plazas}, {Prat}, {Rau}, {Romer}, {Roodman}, {Rooney},
  {Rozo}, {Rykoff}, {Sako}, {Samuroff}, {S{\'a}nchez}, {Sanchez}, {Saro},
  {Scarpine}, {Schubnell}, {Scolnic}, {Serrano}, {Sevilla-Noarbe}, {Sheldon},
  {Smith}, {Smith}, {Suchyta}, {Swanson}, {Tarle}, {Thomas}, {To}, {Troxel},
  {Tucker}, {Varga}, {von der Linden}, {Walker}, {Wechsler}, {Weller},
  {Wilkinson}, {Wu}, {Yanny}, {Zhang}, {Zhang}, {Zuntz}, \& {DES
  Collaboration}}]{abbott20}
{Abbott}, T.~M.~C., {Aguena}, M., {Alarcon}, A., {et~al.} 2020,
  \href{http://dx.doi.org/10.1103/PhysRevD.102.023509}{\color{magenta}\prd},
  \href{https://ui.adsabs.harvard.edu/abs/2020PhRvD.102b3509A}{102, 023509}

\bibitem[{{Allen} {et~al.}(2011){Allen}, {Evrard}, \& {Mantz}}]{allen}
{Allen}, S.~W., {Evrard}, A.~E., \& {Mantz}, A.~B. 2011,
  \href{http://dx.doi.org/10.1146/annurev-astro-081710-102514}{\color{magenta}\araa},
  \href{https://ui.adsabs.harvard.edu/abs/2011ARA&A..49..409A}{49, 409}

\bibitem[{{Amendola} {et~al.}(2018){Amendola}, {Appleby}, {Avgoustidis},
  {Bacon}, {Baker}, {Baldi}, {Bartolo}, {Blanchard}, {Bonvin}, {Borgani},
  {Branchini}, {Burrage}, {Camera}, {Carbone}, {Casarini}, {Cropper}, {de
  Rham}, {Dietrich}, {Di Porto}, {Durrer}, {Ealet}, {Ferreira}, {Finelli},
  {Garc{\'{\i}}a-Bellido}, {Giannantonio}, {Guzzo}, {Heavens}, {Heisenberg},
  {Heymans}, {Hoekstra}, {Hollenstein}, {Holmes}, {Hwang}, {Jahnke},
  {Kitching}, {Koivisto}, {Kunz}, {La Vacca}, {Linder}, {March}, {Marra},
  {Martins}, {Majerotto}, {Markovic}, {Marsh}, {Marulli}, {Massey}, {Mellier},
  {Montanari}, {Mota}, {Nunes}, {Percival}, {Pettorino}, {Porciani},
  {Quercellini}, {Read}, {Rinaldi}, {Sapone}, {Sawicki}, {Scaramella},
  {Skordis}, {Simpson}, {Taylor}, {Thomas}, {Trotta}, {Verde}, {Vernizzi},
  {Vollmer}, {Wang}, {Weller}, \& {Zlosnik}}]{amendola2018}
{Amendola}, L., {Appleby}, S., {Avgoustidis}, A., {et~al.} 2018,
  \href{http://dx.doi.org/10.1007/s41114-017-0010-3}{\color{magenta}Living
  Reviews in Relativity},
  \href{http://adsabs.harvard.edu/abs/2018LRR....21....2A}{21, 2}

\bibitem[{{Angulo} {et~al.}(2012){Angulo}, {Springel}, {White}, {Jenkins},
  {Baugh}, \& {Frenk}}]{angulo2012}
{Angulo}, R.~E., {Springel}, V., {White}, S.~D.~M., {et~al.} 2012,
  \href{http://dx.doi.org/10.1111/j.1365-2966.2012.21830.x}{\color{magenta}\mnras},
  \href{https://ui.adsabs.harvard.edu/abs/2012MNRAS.426.2046A}{426, 2046}

\bibitem[{{Asgari} {et~al.}(2021){Asgari}, {Lin}, {Joachimi}, {Giblin},
  {Heymans}, {Hildebrandt}, {Kannawadi}, {St{\"o}lzner}, {Tr{\"o}ster}, {van
  den Busch}, {Wright}, {Bilicki}, {Blake}, {de Jong}, {Dvornik}, {Erben},
  {Getman}, {Hoekstra}, {K{\"o}hlinger}, {Kuijken}, {Miller}, {Radovich},
  {Schneider}, {Shan}, \& {Valentijn}}]{asgari2020}
{Asgari}, M., {Lin}, C.-A., {Joachimi}, B., {et~al.} 2021,
  \href{http://dx.doi.org/10.1051/0004-6361/202039070}{\color{magenta}\aap},
  \href{https://ui.adsabs.harvard.edu/abs/2021A&A...645A.104A}{645, A104}

\bibitem[{{Balm{\`e}s} {et~al.}(2014){Balm{\`e}s}, {Rasera}, {Corasaniti}, \&
  {Alimi}}]{balmes14}
{Balm{\`e}s}, I., {Rasera}, Y., {Corasaniti}, P.~S., \& {Alimi}, J.~M. 2014,
  \href{http://dx.doi.org/10.1093/mnras/stt2050}{\color{magenta}\mnras},
  \href{https://ui.adsabs.harvard.edu/abs/2014MNRAS.437.2328B}{437, 2328}

\bibitem[{{Bardeau} {et~al.}(2007){Bardeau}, {Soucail}, {Kneib}, {Czoske},
  {Ebeling}, {Hudelot}, {Smail}, \& {Smith}}]{citlens1}
{Bardeau}, S., {Soucail}, G., {Kneib}, J.~P., {et~al.} 2007,
  \href{http://dx.doi.org/10.1051/0004-6361:20077443}{\color{magenta}\aap},
  \href{https://ui.adsabs.harvard.edu/abs/2007A&A...470..449B}{470, 449}

\bibitem[{{Bellagamba} {et~al.}(2018){Bellagamba}, {Roncarelli}, {Maturi}, \&
  {Moscardini}}]{amico}
{Bellagamba}, F., {Roncarelli}, M., {Maturi}, M., \& {Moscardini}, L. 2018,
  \href{http://dx.doi.org/10.1093/mnras/stx2701}{\color{magenta}\mnras},
  \href{https://ui.adsabs.harvard.edu/abs/2018MNRAS.473.5221B}{473, 5221}

\bibitem[{{Bellagamba} {et~al.}(2019){Bellagamba}, {Sereno}, {Roncarelli},
  {Maturi}, {Radovich}, {Bardelli}, {Puddu}, {Moscardini}, {Getman},
  {Hildebrandt}, \& {Napolitano}}]{lensing}
{Bellagamba}, F., {Sereno}, M., {Roncarelli}, M., {et~al.} 2019,
  \href{http://dx.doi.org/10.1093/mnras/stz090}{\color{magenta}\mnras},
  \href{https://ui.adsabs.harvard.edu/abs/2019MNRAS.484.1598B}{484, 1598}

\bibitem[{{Bocquet} {et~al.}(2019){Bocquet}, {Dietrich}, {Schrabback}, {Bleem},
  {Klein}, {Allen}, {Applegate}, {Ashby}, {Bautz}, {Bayliss}, {Benson},
  {Brodwin}, {Bulbul}, {Canning}, {Capasso}, {Carlstrom}, {Chang}, {Chiu},
  {Cho}, {Clocchiatti}, {Crawford}, {Crites}, {de Haan}, {Desai}, {Dobbs},
  {Foley}, {Forman}, {Garmire}, {George}, {Gladders}, {Gonzalez}, {Grandis},
  {Gupta}, {Halverson}, {Hlavacek-Larrondo}, {Hoekstra}, {Holder}, {Holzapfel},
  {Hou}, {Hrubes}, {Huang}, {Jones}, {Khullar}, {Knox}, {Kraft}, {Lee}, {von
  der Linden}, {Luong-Van}, {Mantz}, {Marrone}, {McDonald}, {McMahon}, {Meyer},
  {Mocanu}, {Mohr}, {Morris}, {Padin}, {Patil}, {Pryke}, {Rapetti},
  {Reichardt}, {Rest}, {Ruhl}, {Saliwanchik}, {Saro}, {Sayre}, {Schaffer},
  {Shirokoff}, {Stalder}, {Stanford}, {Staniszewski}, {Stark}, {Story},
  {Strazzullo}, {Stubbs}, {Vanderlinde}, {Vieira}, {Vikhlinin}, {Williamson},
  \& {Zenteno}}]{bocquet19}
{Bocquet}, S., {Dietrich}, J.~P., {Schrabback}, T., {et~al.} 2019,
  \href{http://dx.doi.org/10.3847/1538-4357/ab1f10}{\color{magenta}\apj},
  \href{https://ui.adsabs.harvard.edu/abs/2019ApJ...878...55B}{878, 55}

\bibitem[{{Bocquet} {et~al.}(2016){Bocquet}, {Saro}, {Dolag}, \&
  {Mohr}}]{bocquet}
{Bocquet}, S., {Saro}, A., {Dolag}, K., \& {Mohr}, J.~J. 2016,
  \href{http://dx.doi.org/10.1093/mnras/stv2657}{\color{magenta}\mnras},
  \href{https://ui.adsabs.harvard.edu/abs/2016MNRAS.456.2361B}{456, 2361}

\bibitem[{{B{\"o}hringer} {et~al.}(2004){B{\"o}hringer}, {Schuecker}, {Guzzo},
  {Collins}, {Voges}, {Cruddace}, {Ortiz-Gil}, {Chincarini}, {De Grandi},
  {Edge}, {MacGillivray}, {Neumann}, {Schindler}, \& {Shaver}}]{boringer}
{B{\"o}hringer}, H., {Schuecker}, P., {Guzzo}, L., {et~al.} 2004,
  \href{http://dx.doi.org/10.1051/0004-6361:20034484}{\color{magenta}\aap},
  \href{https://ui.adsabs.harvard.edu/abs/2004A&A...425..367B}{425, 367}

\bibitem[{{Borgani} \& {Kravtsov}(2011)}]{borgani2011}
{Borgani}, S. \& {Kravtsov}, A. 2011,
  \href{http://dx.doi.org/10.1166/asl.2011.1209}{\color{magenta}Advanced
  Science Letters},
  \href{https://ui.adsabs.harvard.edu/abs/2011ASL.....4..204B}{4, 204}

\bibitem[{{Castro} {et~al.}(2021){Castro}, {Borgani}, {Dolag}, {Marra},
  {Quartin}, {Saro}, \& {Sefusatti}}]{castro2020}
{Castro}, T., {Borgani}, S., {Dolag}, K., {et~al.} 2021,
  \href{http://dx.doi.org/10.1093/mnras/staa3473}{\color{magenta}\mnras},
  \href{https://ui.adsabs.harvard.edu/abs/2021MNRAS.500.2316C}{500, 2316}

\bibitem[{{Clerc} {et~al.}(2014){Clerc}, {Adami}, {Lieu}, {Maughan}, {Pacaud},
  {Pierre}, {Sadibekova}, {Smith}, {Valageas}, {Altieri}, {Benoist},
  {Maurogordato}, \& {Willis}}]{clerc}
{Clerc}, N., {Adami}, C., {Lieu}, M., {et~al.} 2014,
  \href{http://dx.doi.org/10.1093/mnras/stu1625}{\color{magenta}\mnras},
  \href{https://ui.adsabs.harvard.edu/abs/2014MNRAS.444.2723C}{444, 2723}

\bibitem[{{Corasaniti} {et~al.}(2021){Corasaniti}, {Sereno}, \&
  {Ettori}}]{corasaniti21}
{Corasaniti}, P.-S., {Sereno}, M., \& {Ettori}, S. 2021,
  \href{http://dx.doi.org/10.3847/1538-4357/abe9a4}{\color{magenta}\apj},
  \href{https://ui.adsabs.harvard.edu/abs/2021ApJ...911...82C}{911, 82}

\bibitem[{{Costanzi} {et~al.}(2019){Costanzi}, {Rozo}, {Simet}, {Zhang},
  {Evrard}, {Mantz}, {Rykoff}, {Jeltema}, {Gruen}, {Allen}, {McClintock},
  {Romer}, {von der Linden}, {Farahi}, {DeRose}, {Varga}, {Weller}, {Giles},
  {Hollowood}, {Bhargava}, {Bermeo-Hernandez}, {Chen}, {Abbott}, {Abdalla},
  {Avila}, {Bechtol}, {Brooks}, {Buckley-Geer}, {Burke}, {Rosell}, {Kind},
  {Carretero}, {Crocce}, {Cunha}, {da Costa}, {Davis}, {De Vicente}, {Diehl},
  {Dietrich}, {Doel}, {Eifler}, {Estrada}, {Flaugher}, {Fosalba}, {Frieman},
  {Garc{\'\i}a-Bellido}, {Gaztanaga}, {Gerdes}, {Giannantonio}, {Gruendl},
  {Gschwend}, {Gutierrez}, {Hartley}, {Honscheid}, {Hoyle}, {James}, {Krause},
  {Kuehn}, {Kuropatkin}, {Lima}, {Lin}, {Maia}, {March}, {Marshall}, {Martini},
  {Menanteau}, {Miller}, {Miquel}, {Mohr}, {Ogando}, {Plazas}, {Roodman},
  {Sanchez}, {Scarpine}, {Schindler}, {Schubnell}, {Serrano}, {Sevilla-Noarbe},
  {Sheldon}, {Smith}, {Soares-Santos}, {Sobreira}, {Suchyta}, {Swanson},
  {Tarle}, {Thomas}, \& {Wechsler}}]{costanzi}
{Costanzi}, M., {Rozo}, E., {Simet}, M., {et~al.} 2019,
  \href{http://dx.doi.org/10.1093/mnras/stz1949}{\color{magenta}\mnras},
  \href{https://ui.adsabs.harvard.edu/abs/2019MNRAS.488.4779C}{488, 4779}

\bibitem[{{Cui} {et~al.}(2012){Cui}, {Borgani}, {Dolag}, {Murante}, \&
  {Tornatore}}]{cui2012}
{Cui}, W., {Borgani}, S., {Dolag}, K., {Murante}, G., \& {Tornatore}, L. 2012,
  \href{http://dx.doi.org/10.1111/j.1365-2966.2012.21037.x}{\color{magenta}\mnras},
  \href{https://ui.adsabs.harvard.edu/abs/2012MNRAS.423.2279C}{423, 2279}

\bibitem[{{Dark Energy Survey Collaboration} {et~al.}(2016){Dark Energy Survey
  Collaboration}, {Abbott}, {Abdalla}, {Aleksi{\'c}}, {Allam}, {Amara},
  {Bacon}, {Balbinot}, {Banerji}, {Bechtol}, {Benoit-L{\'e}vy}, {Bernstein},
  {Bertin}, {Blazek}, {Bonnett}, {Bridle}, {Brooks}, {Brunner}, {Buckley-Geer},
  {Burke}, {Caminha}, {Capozzi}, {Carlsen}, {Carnero-Rosell}, {Carollo},
  {Carrasco-Kind}, {Carretero}, {Castander}, {Clerkin}, {Collett}, {Conselice},
  {Crocce}, {Cunha}, {D'Andrea}, {da Costa}, {Davis}, {Desai}, {Diehl},
  {Dietrich}, {Dodelson}, {Doel}, {Drlica-Wagner}, {Estrada}, {Etherington},
  {Evrard}, {Fabbri}, {Finley}, {Flaugher}, {Foley}, {Fosalba}, {Frieman},
  {Garc{\'\i}a-Bellido}, {Gaztanaga}, {Gerdes}, {Giannantonio}, {Goldstein},
  {Gruen}, {Gruendl}, {Guarnieri}, {Gutierrez}, {Hartley}, {Honscheid}, {Jain},
  {James}, {Jeltema}, {Jouvel}, {Kessler}, {King}, {Kirk}, {Kron}, {Kuehn},
  {Kuropatkin}, {Lahav}, {Li}, {Lima}, {Lin}, {Maia}, {Makler}, {Manera},
  {Maraston}, {Marshall}, {Martini}, {McMahon}, {Melchior}, {Merson}, {Miller},
  {Miquel}, {Mohr}, {Morice-Atkinson}, {Naidoo}, {Neilsen}, {Nichol}, {Nord},
  {Ogando}, {Ostrovski}, {Palmese}, {Papadopoulos}, {Peiris}, {Peoples},
  {Percival}, {Plazas}, {Reed}, {Refregier}, {Romer}, {Roodman}, {Ross},
  {Rozo}, {Rykoff}, {Sadeh}, {Sako}, {S{\'a}nchez}, {Sanchez}, {Santiago},
  {Scarpine}, {Schubnell}, {Sevilla-Noarbe}, {Sheldon}, {Smith}, {Smith},
  {Soares-Santos}, {Sobreira}, {Soumagnac}, {Suchyta}, {Sullivan}, {Swanson},
  {Tarle}, {Thaler}, {Thomas}, {Thomas}, {Tucker}, {Vieira}, {Vikram},
  {Walker}, {Wechsler}, {Weller}, {Wester}, {Whiteway}, {Wilcox}, {Yanny},
  {Zhang}, \& {Zuntz}}]{desCollab}
{Dark Energy Survey Collaboration}, {Abbott}, T., {Abdalla}, F.~B., {et~al.}
  2016, \href{http://dx.doi.org/10.1093/mnras/stw641}{\color{magenta}\mnras},
  \href{https://ui.adsabs.harvard.edu/abs/2016MNRAS.460.1270D}{460, 1270}

\bibitem[{{de Jong} {et~al.}(2015){de Jong}, {Verdoes Kleijn}, {Boxhoorn},
  {Buddelmeijer}, {Capaccioli}, {Getman}, {Grado}, {Helmich}, {Huang},
  {Irisarri}, {Kuijken}, {La Barbera}, {McFarland}, {Napolitano}, {Radovich},
  {Sikkema}, {Valentijn}, {Begeman}, {Brescia}, {Cavuoti}, {Choi}, {Cordes},
  {Covone}, {Dall'Ora}, {Hildebrandt}, {Longo}, {Nakajima}, {Paolillo},
  {Puddu}, {Rifatto}, {Tortora}, {van Uitert}, {Buddendiek},
  {Harnois-D{\'e}raps}, {Erben}, {Eriksen}, {Heymans}, {Hoekstra}, {Joachimi},
  {Kitching}, {Klaes}, {Koopmans}, {K{\"o}hlinger}, {Roy}, {Sif{\'o}n},
  {Schneider}, {Sutherland}, {Viola}, \& {Vriend}}]{dejong2015}
{de Jong}, J. T.~A., {Verdoes Kleijn}, G.~A., {Boxhoorn}, D.~R., {et~al.} 2015,
  \href{http://dx.doi.org/10.1051/0004-6361/201526601}{\color{magenta}\aap},
  \href{https://ui.adsabs.harvard.edu/abs/2015A&A...582A..62D}{582, A62}

\bibitem[{{de Jong} {et~al.}(2017){de Jong}, {Verdoes Kleijn}, {Erben},
  {Hildebrandt}, {Kuijken}, {Sikkema}, {Brescia}, {Bilicki}, {Napolitano},
  {Amaro}, {Begeman}, {Boxhoorn}, {Buddelmeijer}, {Cavuoti}, {Getman}, {Grado},
  {Helmich}, {Huang}, {Irisarri}, {La Barbera}, {Longo}, {McFarland},
  {Nakajima}, {Paolillo}, {Puddu}, {Radovich}, {Rifatto}, {Tortora},
  {Valentijn}, {Vellucci}, {Vriend}, {Amon}, {Blake}, {Choi}, {Conti}, {Gwyn},
  {Herbonnet}, {Heymans}, {Hoekstra}, {Klaes}, {Merten}, {Miller}, {Schneider},
  \& {Viola}}]{dejong}
{de Jong}, J. T.~A., {Verdoes Kleijn}, G.~A., {Erben}, T., {et~al.} 2017,
  \href{http://dx.doi.org/10.1051/0004-6361/201730747}{\color{magenta}\aap},
  \href{https://ui.adsabs.harvard.edu/abs/2017A&A...604A.134D}{604, A134}

\bibitem[{{Despali} {et~al.}(2016){Despali}, {Giocoli}, {Angulo}, {Tormen},
  {Sheth}, {Baso}, \& {Moscardini}}]{despali}
{Despali}, G., {Giocoli}, C., {Angulo}, R.~E., {et~al.} 2016,
  \href{http://dx.doi.org/10.1093/mnras/stv2842}{\color{magenta}\mnras},
  \href{https://ui.adsabs.harvard.edu/abs/2016MNRAS.456.2486D}{456, 2486}

\bibitem[{{Edge} {et~al.}(2013){Edge}, {Sutherland}, {Kuijken}, {Driver},
  {McMahon}, {Eales}, \& {Emerson}}]{edge}
{Edge}, A., {Sutherland}, W., {Kuijken}, K., {et~al.} 2013, The Messenger,
  \href{https://ui.adsabs.harvard.edu/abs/2013Msngr.154...32E}{154, 32}

\bibitem[{{Euclid Collaboration} {et~al.}(2019){Euclid Collaboration}, {Adam},
  {Vannier}, {Maurogordato}, {Biviano}, {Adami}, {Ascaso}, {Bellagamba},
  {Benoist}, {Cappi}, {D{\'\i}az-S{\'a}nchez}, {Durret}, {Farrens}, {Gonzalez},
  {Iovino}, {Licitra}, {Maturi}, {Mei}, {Merson}, {Munari}, {Pell{\'o}},
  {Ricci}, {Rocci}, {Roncarelli}, {Sarron}, {Amoura}, {Andreon}, {Apostolakos},
  {Arnaud}, {Bardelli}, {Bartlett}, {Baugh}, {Borgani}, {Brodwin}, {Castander},
  {Castignani}, {Cucciati}, {De Lucia}, {Dubath}, {Fosalba}, {Giocoli},
  {Hoekstra}, {Mamon}, {Melin}, {Moscardini}, {Paltani}, {Radovich},
  {Sartoris}, {Schultheis}, {Sereno}, {Weller}, {Burigana}, {Carvalho},
  {Corcione}, {Kurki-Suonio}, {Lilje}, {Sirri}, {Toledo-Moreo}, \&
  {Zamorani}}]{adam2019}
{Euclid Collaboration}, {Adam}, R., {Vannier}, M., {et~al.} 2019,
  \href{http://dx.doi.org/10.1051/0004-6361/201935088}{\color{magenta}\aap},
  \href{https://ui.adsabs.harvard.edu/abs/2019A&A...627A..23E}{627, A23}

\bibitem[{{Giocoli} {et~al.}(2021){Giocoli}, {Marulli}, {Moscardini}, {Sereno},
  {Veropalumbo}, {Gigante}, {Maturi}, {Radovich}, {Bellagamba}, {Roncarelli},
  {Bardelli}, {Contarini}, {Covone}, {Harnois-D{\'e}raps}, {Ingoglia}, {Lesci},
  {Nanni}, \& {Puddu}}]{giocoli2020}
{Giocoli}, C., {Marulli}, F., {Moscardini}, L., {et~al.} 2021,
  \href{http://dx.doi.org/10.1051/0004-6361/202140795}{\color{magenta}\aap},
  \href{https://ui.adsabs.harvard.edu/abs/2021A&A...653A..19G}{653, A19}

\bibitem[{{Giocoli} {et~al.}(2018){Giocoli}, {Moscardini}, {Baldi},
  {Meneghetti}, \& {Metcalf}}]{giocoliPeaks}
{Giocoli}, C., {Moscardini}, L., {Baldi}, M., {Meneghetti}, M., \& {Metcalf},
  R.~B. 2018,
  \href{http://dx.doi.org/10.1093/mnras/sty1312}{\color{magenta}\mnras},
  \href{https://ui.adsabs.harvard.edu/abs/2018MNRAS.478.5436G}{478, 5436}

\bibitem[{{Giocoli} {et~al.}(2012){Giocoli}, {Tormen}, \& {Sheth}}]{giocoli1}
{Giocoli}, C., {Tormen}, G., \& {Sheth}, R.~K. 2012,
  \href{http://dx.doi.org/10.1111/j.1365-2966.2012.20594.x}{\color{magenta}\mnras},
  \href{https://ui.adsabs.harvard.edu/abs/2012MNRAS.422..185G}{422, 185}

\bibitem[{{Hikage} {et~al.}(2019){Hikage}, {Oguri}, {Hamana}, {More},
  {Mandelbaum}, {Takada}, {K{\"o}hlinger}, {Miyatake}, {Nishizawa}, {Aihara},
  {Armstrong}, {Bosch}, {Coupon}, {Ducout}, {Ho}, {Hsieh}, {Komiyama},
  {Lanusse}, {Leauthaud}, {Lupton}, {Medezinski}, {Mineo}, {Miyama},
  {Miyazaki}, {Murata}, {Murayama}, {Shirasaki}, {Sif{\'o}n}, {Simet},
  {Speagle}, {Spergel}, {Strauss}, {Sugiyama}, {Tanaka}, {Utsumi}, {Wang}, \&
  {Yamada}}]{hsc}
{Hikage}, C., {Oguri}, M., {Hamana}, T., {et~al.} 2019,
  \href{http://dx.doi.org/10.1093/pasj/psz010}{\color{magenta}\pasj},
  \href{https://ui.adsabs.harvard.edu/abs/2019PASJ...71...43H}{71, 43}

\bibitem[{{Hildebrandt} {et~al.}(2017){Hildebrandt}, {Viola}, {Heymans},
  {Joudaki}, {Kuijken}, {Blake}, {Erben}, {Joachimi}, {Klaes}, {Miller},
  {Morrison}, {Nakajima}, {Verdoes Kleijn}, {Amon}, {Choi}, {Covone}, {de
  Jong}, {Dvornik}, {Fenech Conti}, {Grado}, {Harnois-D{\'e}raps}, {Herbonnet},
  {Hoekstra}, {K{\"o}hlinger}, {McFarland}, {Mead}, {Merten}, {Napolitano},
  {Peacock}, {Radovich}, {Schneider}, {Simon}, {Valentijn}, {van den Busch},
  {van Uitert}, \& {Van Waerbeke}}]{hildebr}
{Hildebrandt}, H., {Viola}, M., {Heymans}, C., {et~al.} 2017,
  \href{http://dx.doi.org/10.1093/mnras/stw2805}{\color{magenta}\mnras},
  \href{https://ui.adsabs.harvard.edu/abs/2017MNRAS.465.1454H}{465, 1454}

\bibitem[{{Hilton} {et~al.}(2018){Hilton}, {Hasselfield}, {Sif{\'o}n},
  {Battaglia}, {Aiola}, {Bharadwaj}, {Bond}, {Choi}, {Crichton}, {Datta},
  {Devlin}, {Dunkley}, {D{\"u}nner}, {Gallardo}, {Gralla}, {Hincks}, {Ho},
  {Hubmayr}, {Huffenberger}, {Hughes}, {Koopman}, {Kosowsky}, {Louis},
  {Madhavacheril}, {Marriage}, {Maurin}, {McMahon}, {Miyatake}, {Moodley},
  {N{\ae}ss}, {Nati}, {Newburgh}, {Niemack}, {Oguri}, {Page}, {Partridge},
  {Schmitt}, {Sievers}, {Spergel}, {Staggs}, {Trac}, {van Engelen},
  {Vavagiakis}, \& {Wollack}}]{hilton}
{Hilton}, M., {Hasselfield}, M., {Sif{\'o}n}, C., {et~al.} 2018,
  \href{http://dx.doi.org/10.3847/1538-4365/aaa6cb}{\color{magenta}\apjs},
  \href{https://ui.adsabs.harvard.edu/abs/2018ApJS..235...20H}{235, 20}

\bibitem[{{Hinshaw} {et~al.}(2013){Hinshaw}, {Larson}, {Komatsu}, {Spergel},
  {Bennett}, {Dunkley}, {Nolta}, {Halpern}, {Hill}, {Odegard}, {Page}, {Smith},
  {Weiland}, {Gold}, {Jarosik}, {Kogut}, {Limon}, {Meyer}, {Tucker}, {Wollack},
  \& {Wright}}]{wmap}
{Hinshaw}, G., {Larson}, D., {Komatsu}, E., {et~al.} 2013,
  \href{http://dx.doi.org/10.1088/0067-0049/208/2/19}{\color{magenta}\apjs},
  \href{https://ui.adsabs.harvard.edu/abs/2013ApJS..208...19H}{208, 19}

\bibitem[{{Hoekstra} {et~al.}(2012){Hoekstra}, {Mahdavi}, {Babul}, \&
  {Bildfell}}]{citlens3}
{Hoekstra}, H., {Mahdavi}, A., {Babul}, A., \& {Bildfell}, C. 2012,
  \href{http://dx.doi.org/10.1111/j.1365-2966.2012.22072.x}{\color{magenta}\mnras},
  \href{https://ui.adsabs.harvard.edu/abs/2012MNRAS.427.1298H}{427, 1298}

\bibitem[{{Kuijken}(2011)}]{Kuijken}
{Kuijken}, K. 2011, The Messenger,
  \href{https://ui.adsabs.harvard.edu/abs/2011Msngr.146....8K}{146, 8}

\bibitem[{{Kuijken} {et~al.}(2019){Kuijken}, {Heymans}, {Dvornik},
  {Hildebrandt}, {de Jong}, {Wright}, {Erben}, {Bilicki}, {Giblin}, {Shan},
  {Getman}, {Grado}, {Hoekstra}, {Miller}, {Napolitano}, {Paolilo}, {Radovich},
  {Schneider}, {Sutherland }, {Tewes}, {Tortora}, {Valentijn}, \& {Verdoes
  Kleijn}}]{Kuijken2019}
{Kuijken}, K., {Heymans}, C., {Dvornik}, A., {et~al.} 2019,
  \href{http://dx.doi.org/10.1051/0004-6361/201834918}{\color{magenta}\aap},
  \href{https://ui.adsabs.harvard.edu/abs/2019A&A...625A...2K}{625, A2}

\bibitem[{{Lacasa} \& {Grain}(2019)}]{lacasa}
{Lacasa}, F. \& {Grain}, J. 2019,
  \href{http://dx.doi.org/10.1051/0004-6361/201834343}{\color{magenta}\aap},
  \href{https://ui.adsabs.harvard.edu/abs/2019A&A...624A..61L}{624, A61}

\bibitem[{{Laureijs} {et~al.}(2011){Laureijs}, {Amiaux}, {Arduini},
  {Augu{\`e}res}, {Brinchmann}, {Cole}, {Cropper}, {Dabin}, {Duvet}, {Ealet},
  {Garilli}, {Gondoin}, {Guzzo}, {Hoar}, {Hoekstra}, {Holmes}, {Kitching},
  {Maciaszek}, {Mellier}, {Pasian}, {Percival}, {Rhodes}, {Saavedra Criado},
  {Sauvage}, {Scaramella}, {Valenziano}, {Warren}, {Bender}, {Castander},
  {Cimatti}, {Le F{\`e}vre}, {Kurki-Suonio}, {Levi}, {Lilje}, {Meylan},
  {Nichol}, {Pedersen}, {Popa}, {Rebolo Lopez}, {Rix}, {Rottgering},
  {Zeilinger}, {Grupp}, {Hudelot}, {Massey}, {Meneghetti}, {Miller}, {Paltani},
  {Paulin-Henriksson}, {Pires}, {Saxton}, {Schrabback}, {Seidel}, {Walsh},
  {Aghanim}, {Amendola}, {Bartlett}, {Baccigalupi}, {Beaulieu}, {Benabed},
  {Cuby}, {Elbaz}, {Fosalba}, {Gavazzi}, {Helmi}, {Hook}, {Irwin}, {Kneib},
  {Kunz}, {Mannucci}, {Moscardini}, {Tao}, {Teyssier}, {Weller}, {Zamorani},
  {Zapatero Osorio}, {Boulade}, {Foumond}, {Di Giorgio}, {Guttridge}, {James},
  {Kemp}, {Martignac}, {Spencer}, {Walton}, {Bl{\"u}mchen}, {Bonoli},
  {Bortoletto}, {Cerna}, {Corcione}, {Fabron}, {Jahnke}, {Ligori}, {Madrid},
  {Martin}, {Morgante}, {Pamplona}, {Prieto}, {Riva}, {Toledo}, {Trifoglio},
  {Zerbi}, {Abdalla}, {Douspis}, {Grenet}, {Borgani}, {Bouwens}, {Courbin},
  {Delouis}, {Dubath}, {Fontana}, {Frailis}, {Grazian}, {Koppenh{\"o}fer},
  {Mansutti}, {Melchior}, {Mignoli}, {Mohr}, {Neissner}, {Noddle}, {Poncet},
  {Scodeggio}, {Serrano}, {Shane}, {Starck}, {Surace}, {Taylor},
  {Verdoes-Kleijn}, {Vuerli}, {Williams}, {Zacchei}, {Altieri}, {Escudero
  Sanz}, {Kohley}, {Oosterbroek}, {Astier}, {Bacon}, {Bardelli}, {Baugh},
  {Bellagamba}, {Benoist}, {Bianchi}, {Biviano}, {Branchini}, {Carbone},
  {Cardone}, {Clements}, {Colombi}, {Conselice}, {Cresci}, {Deacon}, {Dunlop},
  {Fedeli}, {Fontanot}, {Franzetti}, {Giocoli}, {Garcia-Bellido}, {Gow},
  {Heavens}, {Hewett}, {Heymans}, {Holland}, {Huang}, {Ilbert}, {Joachimi},
  {Jennins}, {Kerins}, {Kiessling}, {Kirk}, {Kotak}, {Krause}, {Lahav}, {van
  Leeuwen}, {Lesgourgues}, {Lombardi}, {Magliocchetti}, {Maguire}, {Majerotto},
  {Maoli}, {Marulli}, {Maurogordato}, {McCracken}, {McLure}, {Melchiorri},
  {Merson}, {Moresco}, {Nonino}, {Norberg}, {Peacock}, {Pello}, {Penny},
  {Pettorino}, {Di Porto}, {Pozzetti}, {Quercellini}, {Radovich}, {Rassat},
  {Roche}, {Ronayette}, {Rossetti}, {Sartoris}, {Schneider}, {Semboloni},
  {Serjeant}, {Simpson}, {Skordis}, {Smadja}, {Smartt}, {Spano}, {Spiro},
  {Sullivan}, {Tilquin}, {Trotta}, {Verde}, {Wang}, {Williger}, {Zhao},
  {Zoubian}, \& {Zucca}}]{laureijs}
{Laureijs}, R., {Amiaux}, J., {Arduini}, S., {et~al.} 2011,
  \href{https://ui.adsabs.harvard.edu/abs/2011arXiv1110.3193L}{arXiv e-prints,
  arXiv:1110.3193}

\bibitem[{{LSST Dark Energy Science Collaboration}(2012)}]{LSST2012}
{LSST Dark Energy Science Collaboration}. 2012,
  \href{https://ui.adsabs.harvard.edu/abs/2012arXiv1211.0310L}{arXiv e-prints,
  arXiv:1211.0310}

\bibitem[{{Martinet} {et~al.}(2018){Martinet}, {Schneider}, {Hildebrandt},
  {Shan}, {Asgari}, {Dietrich}, {Harnois-D{\'e}raps}, {Erben}, {Grado},
  {Heymans}, {Hoekstra}, {Klaes}, {Kuijken}, {Merten}, \&
  {Nakajima}}]{martinet17}
{Martinet}, N., {Schneider}, P., {Hildebrandt}, H., {et~al.} 2018,
  \href{http://dx.doi.org/10.1093/mnras/stx2793}{\color{magenta}\mnras},
  \href{https://ui.adsabs.harvard.edu/abs/2018MNRAS.474..712M}{474, 712}

\bibitem[{{Marulli} {et~al.}(2021){Marulli}, {Veropalumbo},
  {Garc{\'\i}a-Farieta}, {Moresco}, {Moscardini}, \& {Cimatti}}]{marulli2020}
{Marulli}, F., {Veropalumbo}, A., {Garc{\'\i}a-Farieta}, J.~E., {et~al.} 2021,
  \href{http://dx.doi.org/10.3847/1538-4357/ac0e8c}{\color{magenta}\apj},
  \href{https://ui.adsabs.harvard.edu/abs/2021ApJ...920...13M}{920, 13}

\bibitem[{{Marulli} {et~al.}(2016){Marulli}, {Veropalumbo}, \& {Moresco}}]{cbl}
{Marulli}, F., {Veropalumbo}, A., \& {Moresco}, M. 2016,
  \href{http://dx.doi.org/10.1016/j.ascom.2016.01.005}{\color{magenta}Astronomy
  and Computing},
  \href{https://ui.adsabs.harvard.edu/abs/2016A&C....14...35M}{14, 35}

\bibitem[{{Marulli} {et~al.}(2017){Marulli}, {Veropalumbo}, {Moscardini},
  {Cimatti}, \& {Dolag}}]{marulli2017}
{Marulli}, F., {Veropalumbo}, A., {Moscardini}, L., {Cimatti}, A., \& {Dolag},
  K. 2017,
  \href{http://dx.doi.org/10.1051/0004-6361/201526885}{\color{magenta}\aap},
  \href{http://adsabs.harvard.edu/abs/2017A%26A...599A.106M}{599, A106}

\bibitem[{{Marulli} {et~al.}(2018){Marulli}, {Veropalumbo}, {Sereno},
  {Moscardini}, {Pacaud}, {Pierre}, {Plionis}, {Cappi}, {Adami}, {Alis},
  {Altieri}, {Birkinshaw}, {Ettori}, {Faccioli}, {Gastaldello}, {Koulouridis},
  {Lidman}, {Le F{\`e}vre}, {Maurogordato}, {Poggianti}, {Pompei},
  {Sadibekova}, \& {Valtchanov}}]{marulli2018}
{Marulli}, F., {Veropalumbo}, A., {Sereno}, M., {et~al.} 2018,
  \href{http://dx.doi.org/10.1051/0004-6361/201833238}{\color{magenta}\aap},
  \href{https://ui.adsabs.harvard.edu/abs/2018A%26A...620A...1M}{620, A1}

\bibitem[{{Maturi} {et~al.}(2019){Maturi}, {Bellagamba}, {Radovich},
  {Roncarelli}, {Sereno}, {Moscardini}, {Bardelli}, \& {Puddu}}]{catalogue}
{Maturi}, M., {Bellagamba}, F., {Radovich}, M., {et~al.} 2019,
  \href{http://dx.doi.org/10.1093/mnras/stz294}{\color{magenta}\mnras},
  \href{https://ui.adsabs.harvard.edu/abs/2019MNRAS.485..498M}{485, 498}

\bibitem[{{Maturi} {et~al.}(2011){Maturi}, {Fedeli}, \& {Moscardini}}]{peaks2}
{Maturi}, M., {Fedeli}, C., \& {Moscardini}, L. 2011,
  \href{http://dx.doi.org/10.1111/j.1365-2966.2011.18958.x}{\color{magenta}\mnras},
  \href{https://ui.adsabs.harvard.edu/abs/2011MNRAS.416.2527M}{416, 2527}

\bibitem[{{Maturi} {et~al.}(2005){Maturi}, {Meneghetti}, {Bartelmann}, {Dolag},
  \& {Moscardini}}]{maturicit}
{Maturi}, M., {Meneghetti}, M., {Bartelmann}, M., {Dolag}, K., \& {Moscardini},
  L. 2005,
  \href{http://dx.doi.org/10.1051/0004-6361:20042600}{\color{magenta}\aap},
  \href{https://ui.adsabs.harvard.edu/abs/2005A&A...442..851M}{442, 851}

\bibitem[{{Melchior} {et~al.}(2015){Melchior}, {Suchyta}, {Huff}, {Hirsch},
  {Kacprzak}, {Rykoff}, {Gruen}, {Armstrong}, {Bacon}, {Bechtol}, {Bernstein},
  {Bridle}, {Clampitt}, {Honscheid}, {Jain}, {Jouvel}, {Krause}, {Lin},
  {MacCrann}, {Patton}, {Plazas}, {Rowe}, {Vikram}, {Wilcox}, {Young}, {Zuntz},
  {Abbott}, {Abdalla}, {Allam}, {Banerji}, {Bernstein}, {Bernstein}, {Bertin},
  {Buckley-Geer}, {Burke}, {Castander}, {da Costa}, {Cunha}, {Depoy}, {Desai},
  {Diehl}, {Doel}, {Estrada}, {Evrard}, {Neto}, {Fernandez}, {Finley},
  {Flaugher}, {Frieman}, {Gaztanaga}, {Gerdes}, {Gruendl}, {Gutierrez},
  {Jarvis}, {Karliner}, {Kent}, {Kuehn}, {Kuropatkin}, {Lahav}, {Maia},
  {Makler}, {Marriner}, {Marshall}, {Merritt}, {Miller}, {Miquel}, {Mohr},
  {Neilsen}, {Nichol}, {Nord}, {Reil}, {Roe}, {Roodman}, {Sako}, {Sanchez},
  {Santiago}, {Schindler}, {Schubnell}, {Sevilla-Noarbe}, {Sheldon}, {Smith},
  {Soares-Santos}, {Swanson}, {Sypniewski}, {Tarle}, {Thaler}, {Thomas},
  {Tucker}, {Walker}, {Wechsler}, {Weller}, \& {Wester}}]{citlens4}
{Melchior}, P., {Suchyta}, E., {Huff}, E., {et~al.} 2015,
  \href{http://dx.doi.org/10.1093/mnras/stv398}{\color{magenta}\mnras},
  \href{https://ui.adsabs.harvard.edu/abs/2015MNRAS.449.2219M}{449, 2219}

\bibitem[{{Miyazaki} {et~al.}(2018){Miyazaki}, {Oguri}, {Hamana}, {Shirasaki},
  {Koike}, {Komiyama}, {Umetsu}, {Utsumi}, {Okabe}, {More}, {Medezinski},
  {Lin}, {Miyatake}, {Murayama}, {Ota}, \& {Mitsuishi}}]{satoshi}
{Miyazaki}, S., {Oguri}, M., {Hamana}, T., {et~al.} 2018,
  \href{http://dx.doi.org/10.1093/pasj/psx120}{\color{magenta}\pasj},
  \href{https://ui.adsabs.harvard.edu/abs/2018PASJ...70S..27M}{70, S27}

\bibitem[{{Murata} {et~al.}(2019){Murata}, {Oguri}, {Nishimichi}, {Takada},
  {Mandelbaum}, {More}, {Shirasaki}, {Nishizawa}, \& {Osato}}]{murata19}
{Murata}, R., {Oguri}, M., {Nishimichi}, T., {et~al.} 2019,
  \href{http://dx.doi.org/10.1093/pasj/psz092}{\color{magenta}\pasj},
  \href{https://ui.adsabs.harvard.edu/abs/2019PASJ...71..107M}{71, 107}

\bibitem[{{Nanni} {et~al.}(in prep.)}]{nanni2020}
{Nanni}, L. {et~al.} in prep.

\bibitem[{{Okabe} {et~al.}(2010){Okabe}, {Zhang}, {Finoguenov}, {Takada},
  {Smith}, {Umetsu}, \& {Futamase}}]{citlens2}
{Okabe}, N., {Zhang}, Y.~Y., {Finoguenov}, A., {et~al.} 2010,
  \href{http://dx.doi.org/10.1088/0004-637X/721/1/875}{\color{magenta}\apj},
  \href{https://ui.adsabs.harvard.edu/abs/2010ApJ...721..875O}{721, 875}

\bibitem[{{Pacaud} {et~al.}(2018){Pacaud}, {Pierre}, {Melin}, {Adami},
  {Evrard}, {Galli}, {Gastaldello}, {Maughan}, {Sereno}, {Alis}, {Altieri},
  {Birkinshaw}, {Chiappetti}, {Faccioli}, {Giles}, {Horellou}, {Iovino},
  {Koulouridis}, {Le F{\`e}vre}, {Lidman}, {Lieu}, {Maurogordato},
  {Moscardini}, {Plionis}, {Poggianti}, {Pompei}, {Sadibekova}, {Valtchanov},
  \& {Willis}}]{pacaud}
{Pacaud}, F., {Pierre}, M., {Melin}, J.~B., {et~al.} 2018,
  \href{http://dx.doi.org/10.1051/0004-6361/201834022}{\color{magenta}\aap},
  \href{https://ui.adsabs.harvard.edu/abs/2018A&A...620A..10P}{620, A10}

\bibitem[{{Pierre} {et~al.}(2016){Pierre}, {Pacaud}, {Adami}, {Alis},
  {Altieri}, {Baran}, {Benoist}, {Birkinshaw}, {Bongiorno}, {Bremer}, {Brusa},
  {Butler}, {Ciliegi}, {Chiappetti}, {Clerc}, {Corasaniti}, {Coupon}, {De
  Breuck}, {Democles}, {Desai}, {Delhaize}, {Devriendt}, {Dubois}, {Eckert},
  {Elyiv}, {Ettori}, {Evrard}, {Faccioli}, {Farahi}, {Ferrari}, {Finet},
  {Fotopoulou}, {Fourmanoit}, {Gandhi}, {Gastaldello}, {Gastaud},
  {Georgantopoulos}, {Giles}, {Guennou}, {Guglielmo}, {Horellou}, {Husband},
  {Huynh}, {Iovino}, {Kilbinger}, {Koulouridis}, {Lavoie}, {Le Brun}, {Le
  Fevre}, {Lidman}, {Lieu}, {Lin}, {Mantz}, {Maughan}, {Maurogordato},
  {McCarthy}, {McGee}, {Melin}, {Melnyk}, {Menanteau}, {Novak}, {Paltani},
  {Plionis}, {Poggianti}, {Pomarede}, {Pompei}, {Ponman}, {Ramos-Ceja},
  {Ranalli}, {Rapetti}, {Raychaudury}, {Reiprich}, {Rottgering}, {Rozo},
  {Rykoff}, {Sadibekova}, {Santos}, {Sauvageot}, {Schimd}, {Sereno}, {Smith},
  {Smol{\v c}i{\'c}}, {Snowden}, {Spergel}, {Stanford}, {Surdej}, {Valageas},
  {Valotti}, {Valtchanov}, {Vignali}, {Willis}, \& {Ziparo}}]{pierre2016}
{Pierre}, M., {Pacaud}, F., {Adami}, C., {et~al.} 2016,
  \href{http://dx.doi.org/10.1051/0004-6361/201526766}{\color{magenta}\aap},
  \href{http://adsabs.harvard.edu/abs/2016A%26A...592A...1P}{592, A1}

\bibitem[{{Planck Collaboration} {et~al.}(2020){Planck Collaboration},
  {Aghanim}, {Akrami}, {Ashdown}, {Aumont}, {Baccigalupi}, {Ballardini},
  {Banday}, {Barreiro}, {Bartolo}, {Basak}, {Battye}, {Benabed}, {Bernard},
  {Bersanelli}, {Bielewicz}, {Bock}, {Bond}, {Borrill}, {Bouchet}, {Boulanger},
  {Bucher}, {Burigana}, {Butler}, {Calabrese}, {Cardoso}, {Carron},
  {Challinor}, {Chiang}, {Chluba}, {Colombo}, {Combet}, {Contreras}, {Crill},
  {Cuttaia}, {de Bernardis}, {de Zotti}, {Delabrouille}, {Delouis}, {Di
  Valentino}, {Diego}, {Dor{\'e}}, {Douspis}, {Ducout}, {Dupac}, {Dusini},
  {Efstathiou}, {Elsner}, {En{\ss}lin}, {Eriksen}, {Fantaye}, {Farhang},
  {Fergusson}, {Fernandez-Cobos}, {Finelli}, {Forastieri}, {Frailis},
  {Fraisse}, {Franceschi}, {Frolov}, {Galeotta}, {Galli}, {Ganga},
  {G{\'e}nova-Santos}, {Gerbino}, {Ghosh}, {Gonz{\'a}lez-Nuevo}, {G{\'o}rski},
  {Gratton}, {Gruppuso}, {Gudmundsson}, {Hamann}, {Handley}, {Hansen},
  {Herranz}, {Hildebrandt}, {Hivon}, {Huang}, {Jaffe}, {Jones}, {Karakci},
  {Keih{\"a}nen}, {Keskitalo}, {Kiiveri}, {Kim}, {Kisner}, {Knox},
  {Krachmalnicoff}, {Kunz}, {Kurki-Suonio}, {Lagache}, {Lamarre}, {Lasenby},
  {Lattanzi}, {Lawrence}, {Le Jeune}, {Lemos}, {Lesgourgues}, {Levrier},
  {Lewis}, {Liguori}, {Lilje}, {Lilley}, {Lindholm}, {L{\'o}pez-Caniego},
  {Lubin}, {Ma}, {Mac{\'\i}as-P{\'e}rez}, {Maggio}, {Maino}, {Mandolesi},
  {Mangilli}, {Marcos-Caballero}, {Maris}, {Martin}, {Martinelli},
  {Mart{\'\i}nez-Gonz{\'a}lez}, {Matarrese}, {Mauri}, {McEwen}, {Meinhold},
  {Melchiorri}, {Mennella}, {Migliaccio}, {Millea}, {Mitra},
  {Miville-Desch{\^e}nes}, {Molinari}, {Montier}, {Morgante}, {Moss}, {Natoli},
  {N{\o}rgaard-Nielsen}, {Pagano}, {Paoletti}, {Partridge}, {Patanchon},
  {Peiris}, {Perrotta}, {Pettorino}, {Piacentini}, {Polastri}, {Polenta},
  {Puget}, {Rachen}, {Reinecke}, {Remazeilles}, {Renzi}, {Rocha}, {Rosset},
  {Roudier}, {Rubi{\~n}o-Mart{\'\i}n}, {Ruiz-Granados}, {Salvati}, {Sandri},
  {Savelainen}, {Scott}, {Shellard}, {Sirignano}, {Sirri}, {Spencer},
  {Sunyaev}, {Suur-Uski}, {Tauber}, {Tavagnacco}, {Tenti}, {Toffolatti},
  {Tomasi}, {Trombetti}, {Valenziano}, {Valiviita}, {Van Tent}, {Vibert},
  {Vielva}, {Villa}, {Vittorio}, {Wandelt}, {Wehus}, {White}, {White},
  {Zacchei}, \& {Zonca}}]{planck}
{Planck Collaboration}, {Aghanim}, N., {Akrami}, Y., {et~al.} 2020,
  \href{http://dx.doi.org/10.1051/0004-6361/201833910}{\color{magenta}\aap},
  \href{https://ui.adsabs.harvard.edu/abs/2020A&A...641A...6P}{641, A6}

\bibitem[{{Reischke} {et~al.}(2016){Reischke}, {Maturi}, \&
  {Bartelmann}}]{peaks1}
{Reischke}, R., {Maturi}, M., \& {Bartelmann}, M. 2016,
  \href{http://dx.doi.org/10.1093/mnras/stv2677}{\color{magenta}\mnras},
  \href{https://ui.adsabs.harvard.edu/abs/2016MNRAS.456..641R}{456, 641}

\bibitem[{{Rykoff} {et~al.}(2014){Rykoff}, {Rozo}, {Busha}, {Cunha},
  {Finoguenov}, {Evrard}, {Hao}, {Koester}, {Leauthaud}, {Nord}, {Pierre},
  {Reddick}, {Sadibekova}, {Sheldon}, \& {Wechsler}}]{redmapper}
{Rykoff}, E.~S., {Rozo}, E., {Busha}, M.~T., {et~al.} 2014,
  \href{http://dx.doi.org/10.1088/0004-637X/785/2/104}{\color{magenta}\apj},
  \href{https://ui.adsabs.harvard.edu/abs/2014ApJ...785..104R}{785, 104}

\bibitem[{{Sartoris} {et~al.}(2016){Sartoris}, {Biviano}, {Fedeli}, {Bartlett},
  {Borgani}, {Costanzi}, {Giocoli}, {Moscardini}, {Weller}, {Ascaso},
  {Bardelli}, {Maurogordato}, \& {Viana}}]{sartoriseuclid}
{Sartoris}, B., {Biviano}, A., {Fedeli}, C., {et~al.} 2016,
  \href{http://dx.doi.org/10.1093/mnras/stw630}{\color{magenta}\mnras},
  \href{https://ui.adsabs.harvard.edu/abs/2016MNRAS.459.1764S}{459, 1764}

\bibitem[{{Schrabback} {et~al.}(2018){Schrabback}, {Applegate}, {Dietrich},
  {Hoekstra}, {Bocquet}, {Gonzalez}, {von der Linden}, {McDonald}, {Morrison},
  {Raihan}, {Allen}, {Bayliss}, {Benson}, {Bleem}, {Chiu}, {Desai}, {Foley},
  {de Haan}, {High}, {Hilbert}, {Mantz}, {Massey}, {Mohr}, {Reichardt}, {Saro},
  {Simon}, {Stern}, {Stubbs}, \& {Zenteno}}]{citlens5}
{Schrabback}, T., {Applegate}, D., {Dietrich}, J.~P., {et~al.} 2018,
  \href{http://dx.doi.org/10.1093/mnras/stx2666}{\color{magenta}\mnras},
  \href{https://ui.adsabs.harvard.edu/abs/2018MNRAS.474.2635S}{474, 2635}

\bibitem[{{Sereno}(2015)}]{rescaling}
{Sereno}, M. 2015,
  \href{http://dx.doi.org/10.1093/mnras/stu2505}{\color{magenta}\mnras},
  \href{https://ui.adsabs.harvard.edu/abs/2015MNRAS.450.3665S}{450, 3665}

\bibitem[{{Sereno} \& {Ettori}(2015)}]{serenoettori}
{Sereno}, M. \& {Ettori}, S. 2015,
  \href{http://dx.doi.org/10.1038/ncomms8211}{\color{magenta}Nature
  Communications},
  \href{https://ui.adsabs.harvard.edu/abs/2015NatCo...6.7211.}{6, 7211}

\bibitem[{{Sereno} {et~al.}(2020){Sereno}, {Ettori}, {Lesci}, {Marulli},
  {Maturi}, {Moscardini}, {Radovich}, {Bellagamba}, \&
  {Roncarelli}}]{sereno2020}
{Sereno}, M., {Ettori}, S., {Lesci}, G.~F., {et~al.} 2020,
  \href{http://dx.doi.org/10.1093/mnras/staa1902}{\color{magenta}\mnras},
  \href{https://ui.adsabs.harvard.edu/abs/2020MNRAS.497..894S}{497, 894}

\bibitem[{{Sereno} {et~al.}(2015){Sereno}, {Veropalumbo}, {Marulli}, {Covone},
  {Moscardini}, \& {Cimatti}}]{sereno2015}
{Sereno}, M., {Veropalumbo}, A., {Marulli}, F., {et~al.} 2015,
  \href{http://dx.doi.org/10.1093/mnras/stv280}{\color{magenta}\mnras},
  \href{http://adsabs.harvard.edu/abs/2015MNRAS.449.4147S}{449, 4147}

\bibitem[{{Shan} {et~al.}(2018){Shan}, {Liu}, {Hildebrandt}, {Pan}, {Martinet},
  {Fan}, {Schneider}, {Asgari}, {Harnois-D{\'e}raps}, {Hoekstra}, {Wright},
  {Dietrich}, {Erben}, {Getman}, {Grado}, {Heymans}, {Klaes}, {Kuijken},
  {Merten}, {Puddu}, {Radovich}, \& {Wang}}]{shan17}
{Shan}, H., {Liu}, X., {Hildebrandt}, H., {et~al.} 2018,
  \href{http://dx.doi.org/10.1093/mnras/stx2837}{\color{magenta}\mnras},
  \href{https://ui.adsabs.harvard.edu/abs/2018MNRAS.474.1116S}{474, 1116}

\bibitem[{{Sheth} \& {Tormen}(1999)}]{shethtormen}
{Sheth}, R.~K. \& {Tormen}, G. 1999,
  \href{http://dx.doi.org/10.1046/j.1365-8711.1999.02692.x}{\color{magenta}\mnras},
  \href{https://ui.adsabs.harvard.edu/abs/1999MNRAS.308..119S}{308, 119}

\bibitem[{{Stern} {et~al.}(2019){Stern}, {Dietrich}, {Bocquet}, {Applegate},
  {Mohr}, {Bridle}, {Carrasco Kind}, {Gruen}, {Jarvis}, {Kacprzak}, {Saro},
  {Sheldon}, {Troxel}, {Zuntz}, {Benson}, {Capasso}, {Chiu}, {Desai},
  {Rapetti}, {Reichardt}, {Saliwanchik}, {Schrabback}, {Gupta}, {Abbott},
  {Abdalla}, {Avila}, {Bertin}, {Brooks}, {Burke}, {Carnero Rosell},
  {Carretero}, {Castander}, {D'Andrea}, {da Costa}, {Davis}, {De Vicente},
  {Diehl}, {Doel}, {Estrada}, {Evrard}, {Flaugher}, {Fosalba}, {Frieman},
  {Garc{\'\i}a-Bellido}, {Gaztanaga}, {Gruendl}, {Gschwend}, {Gutierrez},
  {Hollowood}, {Jeltema}, {Kirk}, {Kuehn}, {Kuropatkin}, {Lahav}, {Lima},
  {Maia}, {March}, {Melchior}, {Menanteau}, {Miquel}, {Plazas}, {Romer},
  {Sanchez}, {Schindler}, {Schubnell}, {Sevilla-Noarbe}, {Smith}, {Smith},
  {Sobreira}, {Suchyta}, {Swanson}, {Tarle}, {Walker}, {DES Collaboration}, \&
  {SPT Collaboration}}]{citlens6}
{Stern}, C., {Dietrich}, J.~P., {Bocquet}, S., {et~al.} 2019,
  \href{http://dx.doi.org/10.1093/mnras/stz234}{\color{magenta}\mnras},
  \href{https://ui.adsabs.harvard.edu/abs/2019MNRAS.485...69S}{485, 69}

\bibitem[{{Tinker} {et~al.}(2008){Tinker}, {Kravtsov}, {Klypin}, {Abazajian},
  {Warren}, {Yepes}, {Gottl{\"o}ber}, \& {Holz}}]{tinker}
{Tinker}, J., {Kravtsov}, A.~V., {Klypin}, A., {et~al.} 2008,
  \href{http://dx.doi.org/10.1086/591439}{\color{magenta}\apj},
  \href{https://ui.adsabs.harvard.edu/abs/2008ApJ...688..709T}{688, 709}

\bibitem[{{Troxel} {et~al.}(2018){Troxel}, {MacCrann}, {Zuntz}, {Eifler},
  {Krause}, {Dodelson}, {Gruen}, {Blazek}, {Friedrich}, {Samuroff}, {Prat},
  {Secco}, {Davis}, {Fert{\'e}}, {DeRose}, {Alarcon}, {Amara}, {Baxter},
  {Becker}, {Bernstein}, {Bridle}, {Cawthon}, {Chang}, {Choi}, {De Vicente},
  {Drlica-Wagner}, {Elvin-Poole}, {Frieman}, {Gatti}, {Hartley}, {Honscheid},
  {Hoyle}, {Huff}, {Huterer}, {Jain}, {Jarvis}, {Kacprzak}, {Kirk}, {Kokron},
  {Krawiec}, {Lahav}, {Liddle}, {Peacock}, {Rau}, {Refregier}, {Rollins},
  {Rozo}, {Rykoff}, {S{\'a}nchez}, {Sevilla-Noarbe}, {Sheldon}, {Stebbins},
  {Varga}, {Vielzeuf}, {Wang}, {Wechsler}, {Yanny}, {Abbott}, {Abdalla},
  {Allam}, {Annis}, {Bechtol}, {Benoit-L{\'e}vy}, {Bertin}, {Brooks},
  {Buckley-Geer}, {Burke}, {Carnero Rosell}, {Carrasco Kind}, {Carretero},
  {Castander}, {Crocce}, {Cunha}, {D'Andrea}, {da Costa}, {DePoy}, {Desai},
  {Diehl}, {Dietrich}, {Doel}, {Fernandez}, {Flaugher}, {Fosalba},
  {Garc{\'\i}a-Bellido}, {Gaztanaga}, {Gerdes}, {Giannantonio}, {Goldstein},
  {Gruendl}, {Gschwend}, {Gutierrez}, {James}, {Jeltema}, {Johnson}, {Johnson},
  {Kent}, {Kuehn}, {Kuhlmann}, {Kuropatkin}, {Li}, {Lima}, {Lin}, {Maia},
  {March}, {Marshall}, {Martini}, {Melchior}, {Menanteau}, {Miquel}, {Mohr},
  {Neilsen}, {Nichol}, {Nord}, {Petravick}, {Plazas}, {Romer}, {Roodman},
  {Sako}, {Sanchez}, {Scarpine}, {Schindler}, {Schubnell}, {Smith}, {Smith},
  {Soares-Santos}, {Sobreira}, {Suchyta}, {Swanson}, {Tarle}, {Thomas},
  {Tucker}, {Vikram}, {Walker}, {Weller}, {Zhang}, \& {DES
  Collaboration}}]{troxel}
{Troxel}, M.~A., {MacCrann}, N., {Zuntz}, J., {et~al.} 2018,
  \href{http://dx.doi.org/10.1103/PhysRevD.98.043528}{\color{magenta}\prd},
  \href{https://ui.adsabs.harvard.edu/abs/2018PhRvD..98d3528T}{98, 043528}

\bibitem[{{Velliscig} {et~al.}(2014){Velliscig}, {van Daalen}, {Schaye},
  {McCarthy}, {Cacciato}, {Le Brun}, \& {Dalla Vecchia}}]{velliscig}
{Velliscig}, M., {van Daalen}, M.~P., {Schaye}, J., {et~al.} 2014,
  \href{http://dx.doi.org/10.1093/mnras/stu1044}{\color{magenta}\mnras},
  \href{https://ui.adsabs.harvard.edu/abs/2014MNRAS.442.2641V}{442, 2641}

\bibitem[{{Veropalumbo} {et~al.}(2014){Veropalumbo}, {Marulli}, {Moscardini},
  {Moresco}, \& {Cimatti}}]{veropalumbo2014}
{Veropalumbo}, A., {Marulli}, F., {Moscardini}, L., {Moresco}, M., \&
  {Cimatti}, A. 2014,
  \href{http://dx.doi.org/10.1093/mnras/stu1050}{\color{magenta}\mnras},
  \href{http://adsabs.harvard.edu/abs/2014MNRAS.442.3275V}{442, 3275}

\bibitem[{{Veropalumbo} {et~al.}(2016){Veropalumbo}, {Marulli}, {Moscardini},
  {Moresco}, \& {Cimatti}}]{veropalumbo2016}
{Veropalumbo}, A., {Marulli}, F., {Moscardini}, L., {Moresco}, M., \&
  {Cimatti}, A. 2016,
  \href{http://dx.doi.org/10.1093/mnras/stw306}{\color{magenta}\mnras},
  \href{http://adsabs.harvard.edu/abs/2016MNRAS.458.1909V}{458, 1909}

\bibitem[{{Vikhlinin} {et~al.}(2009){Vikhlinin}, {Kravtsov}, {Burenin},
  {Ebeling}, {Forman}, {Hornstrup}, {Jones}, {Murray}, {Nagai}, {Quintana}, \&
  {Voevodkin}}]{vikhlinin2009}
{Vikhlinin}, A., {Kravtsov}, A.~V., {Burenin}, R.~A., {et~al.} 2009,
  \href{http://dx.doi.org/10.1088/0004-637X/692/2/1060}{\color{magenta}\apj},
  \href{http://adsabs.harvard.edu/abs/2009ApJ...692.1060V}{692, 1060}

\bibitem[{{Watson} {et~al.}(2013){Watson}, {Iliev}, {D'Aloisio}, {Knebe},
  {Shapiro}, \& {Yepes}}]{watson}
{Watson}, W.~A., {Iliev}, I.~T., {D'Aloisio}, A., {et~al.} 2013,
  \href{http://dx.doi.org/10.1093/mnras/stt791}{\color{magenta}\mnras},
  \href{https://ui.adsabs.harvard.edu/abs/2013MNRAS.433.1230W}{433, 1230}

\bibitem[{{Wright} {et~al.}(2019){Wright}, {Hildebrandt}, {Kuijken}, {Erben},
  {Blake}, {Buddelmeijer}, {Choi}, {Cross}, {de Jong}, {Edge},
  {Gonzalez-Fernandez}, {Gonz{\'a}lez Solares}, {Grado}, {Heymans}, {Irwin},
  {Kupcu Yoldas}, {Lewis}, {Mann}, {Napolitano}, {Radovich}, {Schneider},
  {Sif{\'o}n}, {Sutherland}, {Sutorius}, \& {Verdoes Kleijn}}]{wright19}
{Wright}, A.~H., {Hildebrandt}, H., {Kuijken}, K., {et~al.} 2019,
  \href{http://dx.doi.org/10.1051/0004-6361/201834879}{\color{magenta}\aap},
  \href{https://ui.adsabs.harvard.edu/abs/2019A&A...632A..34W}{632, A34}

\end{thebibliography}

\end{document}